\documentclass[aip,cha,reprint]{revtex4-1}

\usepackage{graphicx}
\usepackage{amsmath,amssymb}
\usepackage{hyperref}
\usepackage{rotating}
\usepackage{multirow}
\usepackage{subfigure}
\usepackage{setspace}

\begin{document}

\title{Texture descriptor combining fractal dimension and artificial crawlers}
\author{Wesley Nunes Gon\c{c}alves}
\affiliation{
              Federal University of Mato Grosso do Sul -  Ponta Por\~a - MS, Brazil
              } 

\affiliation{
              email: [wesley.gon\c{c}alves, bruno.brandoli]@ufms.br
              } 

\affiliation{
              S\~{a}o Carlos Institute of Physics (IFSC) - University of S\~{a}o Paulo (USP)\\ S\~ao Carlos - SP, Brazil - 
              http://scg.ifsc.usp.br} 

\author{Bruno Brandoli Machado}
\affiliation{
              Federal University of Mato Grosso do Sul -  Ponta Por\~a - MS, Brazil
              } 

\affiliation{
              email: [wesley.gon\c{c}alves, bruno.brandoli]@ufms.br
              } 

\affiliation{
              S\~{a}o Carlos Institute of Physics (IFSC) - University of S\~{a}o Paulo (USP)\\ S\~ao Carlos - SP, Brazil - 
              http://scg.ifsc.usp.br}

\author{Odemir Martinez Bruno}
\affiliation{
              S\~{a}o Carlos Institute of Physics (IFSC) - University of S\~{a}o Paulo (USP)\\ S\~ao Carlos - SP, Brazil - 
              http://scg.ifsc.usp.br} 
              
\affiliation{
              email: bruno@ifsc.usp.br
              }               

\begin{abstract}
Texture is an important visual attribute used to describe images. There are many methods available for texture analysis. However, they do not capture the details richness of the image surface. In this paper, we propose a new method to describe textures using the artificial crawler model. This model assumes that each agent can interact with the environment and each other. Since this swarm system alone does not achieve a good discrimination, we developed a new method to increase the discriminatory power of artificial crawlers, together with the fractal dimension theory. Here, we estimated the fractal dimension by the Bouligand-Minkowski method due to its precision in quantifying structural properties of images. We validate our method on two texture datasets and the experimental results reveal that our method leads to highly discriminative textural features. The results indicate that our method can be used in different texture applications.


\end{abstract}



\maketitle


\section{Introduction}

The discrimination of visual texture has played an important role in computer vision and image analysis. Although the ability of human beings for texture discrimination is apparently easy, the description by using texture methods has proven to be very complex. Several methods have been proposed to characterize texture images. They are based on \textit{statistical analysis} of the spatial distribution (e.g., co-occurrence matrices \cite{haralickTSMC1973, haralickIEEE1979}, local binary pattern \cite{kashyapPAMI1986} and interaction map \cite{chetverikovPR1998}), \textit{stochastic models} (e.g., Markov random fields \cite{crossPAMI1983,chellappaASSP1985}), \textit{spectral analysis} (e.g., Fourier descriptors \cite{azencottPAMI1997}, Gabor filters \cite{gaborJIEE1946,brandoliACIVS2011} and wavelets transform \cite{mallatPAMI1989,arivazhaganPRL2003}), \textit{structural models} (e.g., mathematical morphology \cite{serra1983} and geometrical analysis \cite{chen1994}), and \textit{complexity analysis} (e.g., 
fractal dimension \cite{mandelbrot1983,odemirIS2008}). Despite the fact they have thoroughly been studied, few methods are able to successfully discriminate the different texture patterns found in nature.

Swarm systems or multi-agent systems, have been long applied in computer vision \cite{liuPAMI1999,wongPR2001,rodinPR2004,guoESA2005,billonACET2008}. In texture analysis, the swarm system can be found in a select group of approaches, such as deterministic the tourist walk \cite{backesPR2010,goncalvesPR2013,goncalves2013}, the ant colony \cite{zhengCEC2003}, and the artificial crawler \cite{ZhangIAT2004,ZhangIJPRAI2005}. The basic idea of the swarm algorithms consists of creating a system by means of the agent interaction, i.e., a distributed agent system with parallel processing, and autonomous computing. In this paper, we propose a novel method for texture analysis based on the artificial crawler model \cite{ZhangIAT2004,ZhangIJPRAI2005}. This swarm system consists of a population of agents, referred here to as artificial crawlers, that interact with each other and the environment, in this case an image. Each artificial crawler occupies a pixel, and its goal is to move to the neighbor pixel of greater 
intensity. The agents store their current position in the image, and a correspondent energy that can wax or wane their lifespan depending on the energy consumption of the image. The population of artificial crawlers stabilizes after a certain number of iterations, i.e., when there is no change in their spatial positions.

In the original swarm system \cite{ZhangIAT2004,ZhangIJPRAI2005} the artificial crawlers only moves in direction of the maximum intensity, thus characterizing regions of high intensities in the image. However, in texture analysis, regions of low intensities are as important as regions of high intensities. Therefore, we propose a new rule of movement that also moves artificial crawler agents in the direction of lower intensity. Our approach differs from the original artificial crawler model in terms of movement: each agent is able to move to the higher altitudes, as well as to lower ones. To quantify the state of the swarm system after the stabilization, we propose to employ the Bouligand-Minkowski fractal dimension method \cite{tricot1995,odemirIS2008}.
The fractal dimension method is widely used to characterize the roughness of a surface, which is related to the physical properties.

We have conducted experiments in two datasets widely accepted in the literature of texture analysis. 
Experimental results have shown that our method overcomes different state-of-the-art methods over Vistex dataset. 
Besides, our approach significantly improves the classification rate compared to the original artificial crawler method.
The superior results rely on two facts: the fractal dimension estimation of the swarm system and the two rules of movement.
On the one hand, the use of both rules of movement characterizes both regions of the texture image.
On the other hand, the fractal dimension improves the ability of discrimination obtained from the swarm system of artificial crawlers.
Moreover, the idea of the fractal dimension estimation can be used for other swarm systems. 

The main contributions of this paper are:
\begin{itemize}
\item a new rule of movement for the artificial crawler method. The original method fails to describe images because it moves the agents to higher intensities only. The proposed method describes images by using two rules of movement, i.e., the swarm system finds the minima and maxima of images.
\item a new methodology to image description based on the energy information acquired from two rules of movement. Although we can find the minima and maxima of images directly, the underlying idea is to characterize the path of movement during the evolution process. In this case, the energy information was considered the most important attribute due to its capacity of representing the interaction between the movement of agents and the environment.
\item to enhance the discriminatory power of our method, we use the energy information and the spatial position of each agent to estimate the fractal dimension of the image surface, in this paper is employed the fractal dimension of Bouligand-Minkowski.
\end{itemize}

This paper is organized as follows.
In Section 2, we describe the artificial crawler model in details.
In Section 3, we present the basis for the fractal dimension and the Bouligand-Minkowski method.
A new method for texture analysis based on fractal dimension of artificial crawlers is presented in Section 4.
Finally, in Section 5 we report the experimental results, followed by the conclusion in Section 6.

\section{Artificial Crawler Model}

The texture method proposed in this study is based on the artificial crawler model proposed in \cite{ZhangIAT2004,ZhangIJPRAI2005}. 
Their agent-based model was first proposed in \cite{ZhangIAT2004} and then extended in \cite{ZhangIJPRAI2005}.
In order to describe this model, let us consider an image which consists of a pair $(\mathcal{I}, I)$ $-$ a finite set $\mathcal{I}$ of pixels and a mapping $I$ that assigns to each pixel $p = (x_p, y_p)$ in $\mathcal{I}$ a intensity $I(p) \in [0, 255]$.
Also, let us consider a neighborhood $\eta(p)$  that consists of pixels $q$ whose Euclidean distance between $p$ and $q$ is smaller or equal to $\sqrt{2}$ (8-connected pixels):
\begin{equation}
 \begin{array}{l}
  \eta(p) = \{q \mid d(p,q) \leq \sqrt{2} \} \\ \\
  d(p,q) = \sqrt{(x_p-x_q)^2 + (y_p-y_q)^2}
 \end{array}
 \label{eq:neighbor}
\end{equation}

\begin{figure}[!htbp]
\centering
\subfigure{\includegraphics[width=0.15\textwidth]{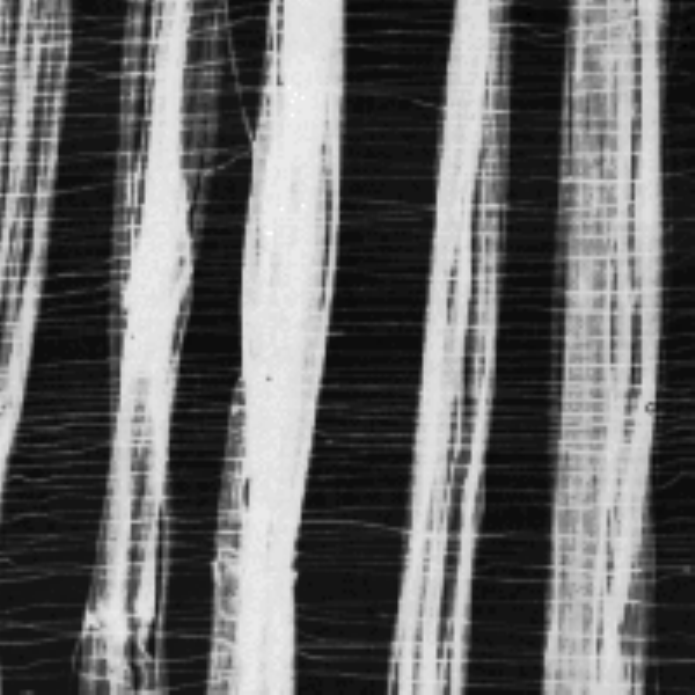}} \\
\subfigure{\includegraphics[width=0.45\textwidth]{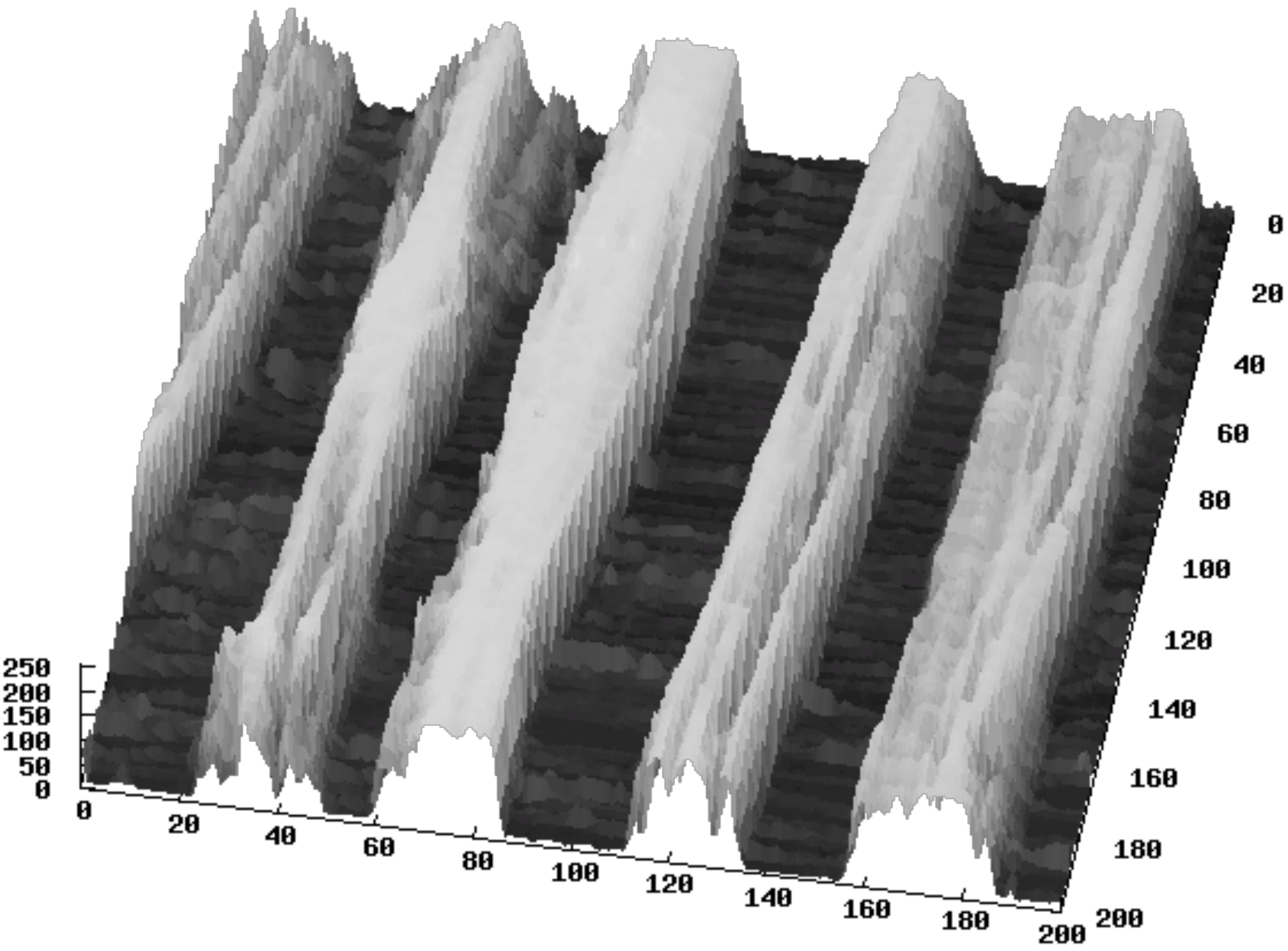}}
\caption{The environment of the artificial crawler. On the top is shown a textured image and below its respective 3D surface.}
\label{fig:environment}
\end{figure}

In image analysis, the artificial crawler model assumes that each agent occupies one pixel of the image.
At each time $t$, artificial crawler $A_t^i = \{ e_t^i, p_t^i \} \, \forall i \in [0,N]$ are characterized by two attributes.
The first attribute $e_t^i$ holds the current level of energy.
Such energy can either wax or wane their lifespan according to energy consumption and influence of the environment.
The second attribute $p_t^i$ is  the current position of the artificial crawler in the image.
The artificial crawlers act upon an environment.
In images, the environment is mapped as a 3D surface with different altitudes that correspond to gray values in z-axis.
Higher intensities pixels supply nutrients to the artificial crawlers (increase its energy), while lower altitudes correspond to the land.
Figure \ref{fig:environment} shows a textured image and the peaks and valleys where the artificial crawlers can increase or decrease its energy live.

The $N$ artificial crawlers begin with equal energy $e_{init}$ and are placed at random on the surface (pixels) of the textured image:
\begin{equation}
\begin{array}{l}
p_0^i = rand(I) \\
e_0^i = e_{init}
\end{array}
\end{equation}

\begin{figure}[!htbp]
\centering
\subfigure[]{\includegraphics[width=0.1\textwidth]{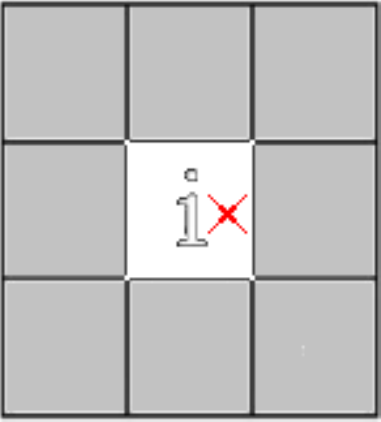}} \hspace{5mm}
\subfigure[]{\includegraphics[width=0.1\textwidth]{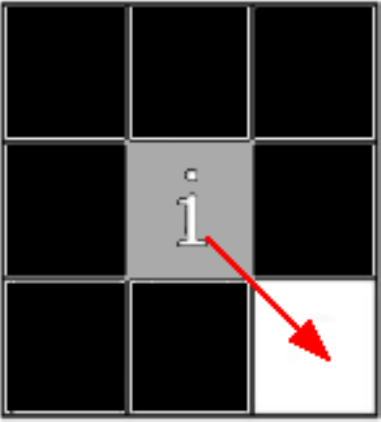}}  \hspace{5mm}
\subfigure[]{\includegraphics[width=0.1\textwidth]{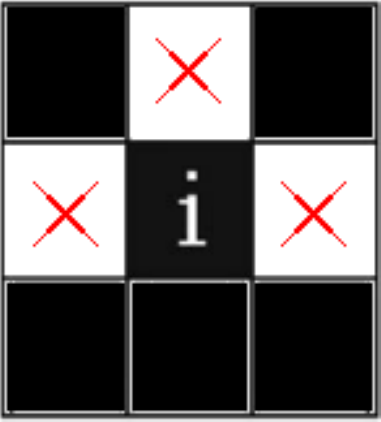}}
\caption{Example of the three possible steps of artificial crawlers considering its 8 neighbors.}
\label{fig:steps}
\end{figure}

Then the evolution process starts following a set of specific rules.
The aim of the artificial crawler is to move to areas of higher altitudes in order to absorb energy and sustain life.
This way, the next step $p_{t+1}^i = f(p_{t}^i)$ depends on the gray level of its neighbors according to Equation \ref{eq:movements}.
First, the artificial crawler settles down if the gray level of its 8 neighbors are lower than itself (Figure \ref{fig:steps} (a)).
Second, the artificial crawler moves to a specific pixel if there exist one of its 8 neighbors with unique higher intensity (Figure \ref{fig:steps} (b)).
Third, if there exist more than one neighbor with higher intensity, an artificial crawler moves to the pixel that was already occupied by another artificial crawler in any time (Figure \ref{fig:steps} (c)).
Otherwise, it moves to one of the pixels randomly.

\begin{equation}
f(p_t^i) = \left\{ \begin{array}{lll}
         p_t^i, & \textrm{if } I(p_t^i) \ge I(p) \, \forall p \in \eta(p_t^i) \\
         p, & \textrm{if } I(p) > I(p_t^i), I(p) > I(q) \, \forall p, q \in \eta(p_t^i), \\
         & p \neq q \\
         p, & \textrm{if } I(p) > I(p_t^i), I(p) \geq  I(q) \, \forall p, q \in \eta(p_t^i), \\
         & p \neq q, p \textrm{ was visited}
         \end{array} \right. 
         \label{eq:movements}
\end{equation}

Given the new position of the artificial crawler, the energy absorption from the environment is performed: 
\begin{equation}
e_{t+1}^i = e_t^i + \lambda I(p_{t+1}^i) - 1
\end{equation}
where $\lambda$ is the rate of absorption over the gray level of the current pixel $I(p_{t+1}^i)$.
All artificial crawlers lose a unit of energy which means that the artificial crawler loses energy at each step if $\lambda*I(p_{t+1}^i) < 1$.
For the default value of $\lambda = 0.01$, it means that the artificial crawler loses energy if it goes to a pixel whose gray level is less than $100$ and gain energy otherwise.
The energy is bounded by limit $e_{max}$, i.e.  if $e_{t+1}^i > e_{max}$ then  $e_{t+1}^i = e_{max}$.
Also, an artificial crawler keep living in the next generation whether its energy is higher than a certain threshold $e_{min}$.

After the energy absorption, the law of the jungle is performed. In this law, an artificial crawler with higher energy eats up another with lower energy if they are in the same pixel, i.e. $A_{t+1}^i$ eats up $A_{t+1}^j$ if $p_{t+1}^i = p_{t+1}^j, e_{t+1}^i \ge e_{t+1}^j, i \neq j$.
This law is inspired in nature and assumes that the artificial crawlers with higher energy are more likely to reach the peaks of the environment.

The evolution process converges to an equilibrium state when no further artificial crawlers are in movement (they are dead or settled down).
In the original method, features are extracted by means of the number of artificial crawlers at each iteration and colonial properties.
Each texture image is represented by four curves of evolution: (1) curve of living artificial crawlers, (2) curve of settled artificial crawlers, (3) curve of colony formation at certain radius and (4) scale distribution of colonies.
This representation has two major drawbacks: (i) the vector obtained is high-dimensional, which lead us to the curse of dimensionality and (ii) the extraction of this vector is very time-consuming due to the colony estimation. 

\section{Fractal Dimension}

In 1977, Mandelbrot introduced a new mathematical concept to model natural phenomena, named fractal geometry \cite{mandelbrot1977}. This formulation received a lot of attention due to its ability to describe irregular shapes and complex objects that Euclidean geometry fails to analyze. In contrast, fractal geometry assumes that an object holds a non-integer dimension. Thus, estimating the fractal dimension of an object is basically related to its complexity. The patterns are characterized in terms of space occupation and self-similarity at different scales. The interactive construction process of the Von Koch curve is a typical example of self-similarity of fractals \cite{mandelbrot1983}.

The first definition of dimension was proposed by the Hausdorff-Besicovitch measure \cite{hausdorffMA1919}, which provided the basis of the fractal dimension theory. He defined a dimension for point sets as a fraction greater than their topological dimension. Formally, given $X \in \Re^{d}$, a geometrical set of points, the Hausdoff-Besicovitch dimension $D_H(X)$ is calculated by:

\begin{equation}
D_H(X) = inf \{s:H^s(X) = 0 \} = sup \{ H^s(X) = \infty \}
\label{eq:hausdorff1}
\end{equation}

\noindent where $H^s(X)$ is the $s$-dimensional Hausdorff measure (in Equation \ref{eq:hausdorff2}).

\begin{equation}
H^s(X) = \lim_{\delta \to 0} inf \left[ \sum_{i=1}^{\infty} |U_i|^s : U_i \textrm{ is a } \delta\textrm{-cover of } X  \right]
\label{eq:hausdorff2}
\end{equation}
where $|.|$ stands for the diameter in $\Re^d$, i.e $|U| = sup|x-y|:x,y \in U$.

In image analysis, the use of the Hausdoff-Besicovitch definition may be impracticable \cite{theilerJOSA1990}. An alternative definition generalized from the topological dimension is commonly used. According to this definition, the fractal dimension D of an object $X$ is:
\begin{equation}
 D(X) = \lim_{\epsilon \to 0} \frac{\log N(\epsilon)}{\log \frac{1}{\epsilon}}
\end{equation}
where $N(\epsilon)$ stands for the number of objects of linear size $\epsilon$ needed to cover the whole object $X$.

There are a lot of algorithms to estimate the fractal dimension of objects or surfaces. The most known algorithms are: box-counting \cite{russellPRL1980}, differential box-counting \cite{chaudhuriPAMI1995}, $\epsilon$-blanket \cite{pelegPAMI1984}, fractal model based on Fractional Brownian motion \cite{pentlandCSCVPR1983}, power spectrum method \cite{pentlandCSCVPR1983}, Bouligand-Minkowski \cite{tricot1995} among others; as well as extensions of fractals, such as multifractals \cite{chaudhariASS2004}, multi-scale fractals \cite{odemirIS2008} and fractal descriptors \cite{BackesCB12,FlorindoBCB12,FlorindoB12,FlorindoSPB13}. 
One of the most accurate methods to estimate the fractal dimension is the Bouligand-Minkowski method \cite{tricot1995,odemirIS2008,backesIJPRAI2009}.
The Boulingand-Minkowski fractal dimension $D_{B}(X)$ depends on a symmetrical structuring element $Y$:

\begin{equation}
\begin{array}{l}
 D_{B}(X,Y) = inf \{\lambda, m_{B}(X,Y,\lambda) = 0 \} \\ \\
 m_{B}(X,Y,\lambda) = \lim_{\epsilon \to 0} \frac{V(\partial X \oplus \epsilon Y)}{\epsilon^{n-\lambda}}
\end{array}
\end{equation}
\noindent where $m_{B}$ is the Bouligand-Minkowski measure, $\epsilon$ is the radius of the  element $Y$ and $V$ is the volume of the dilation between element $Y$ and boundary $\partial X$. To eliminate the explicit dependence on the element $Y$, a simplified version of the Bouligand-Minkowski fractal dimension can be described by using neighborhood techniques as:

\begin{equation}
 D_{B}(X) = \lim_{\epsilon \to 0} \left( D_T - \frac{\log V(X \oplus Y_{\epsilon})}{\log \epsilon} \right)
\end{equation}

For instance considering an object $X \in \Re^3$, the topological dimension $D_T = 3$ and $Y_{\epsilon}$ is a sphere of diameter $\epsilon$. Varying the radius $\epsilon$, it estimates the fractal dimension based on the size of the influence area $V$ created by the dilation of $X$ by $Y_{\epsilon}$.

\section{Proposed Method}
In this section, we describe the proposed method, which is based on the fractal dimension of artificial crawlers.
Basically, our method can be divided into two parts: artificial crawlers are performed in the texture image and then the fractal dimension of these artificial crawlers is estimated. The next sections describe these steps of our method.

\subsection{Artificial Crawler Model in Images}
Although the original artificial crawler method achieves promising results, the idea of moving to pixels with higher intensities does not extract all the richness of textural pattern of the images.
In the method proposed here, the independent artificial crawler is also able to move to lower intensities (valleys).
It allows the model to take full advantage and capture the richness of details present in peaks and valleys of the images.

\begin{figure*}[!htbp]
\centering
\includegraphics[width=0.45\textwidth]{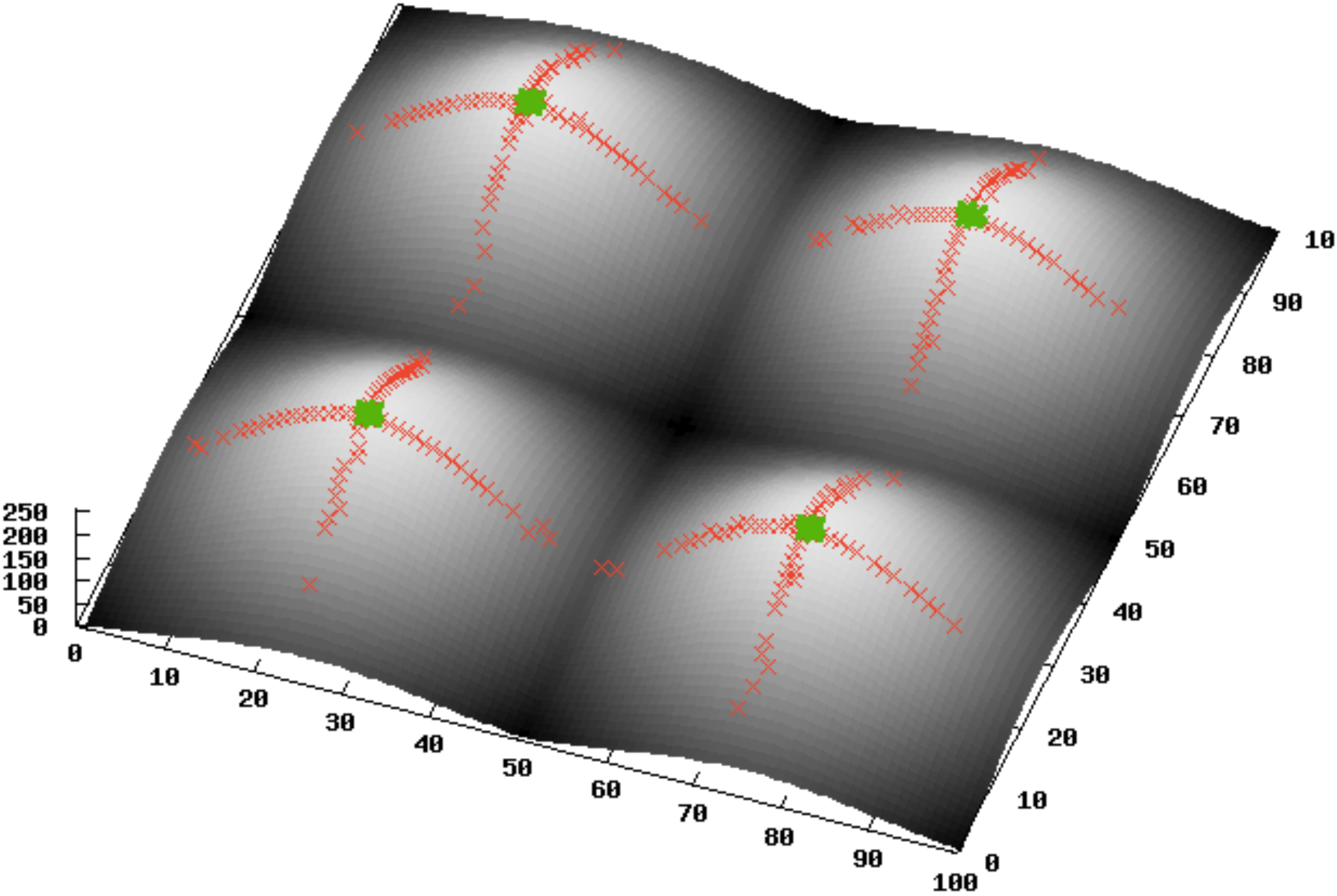}
\includegraphics[width=0.45\textwidth]{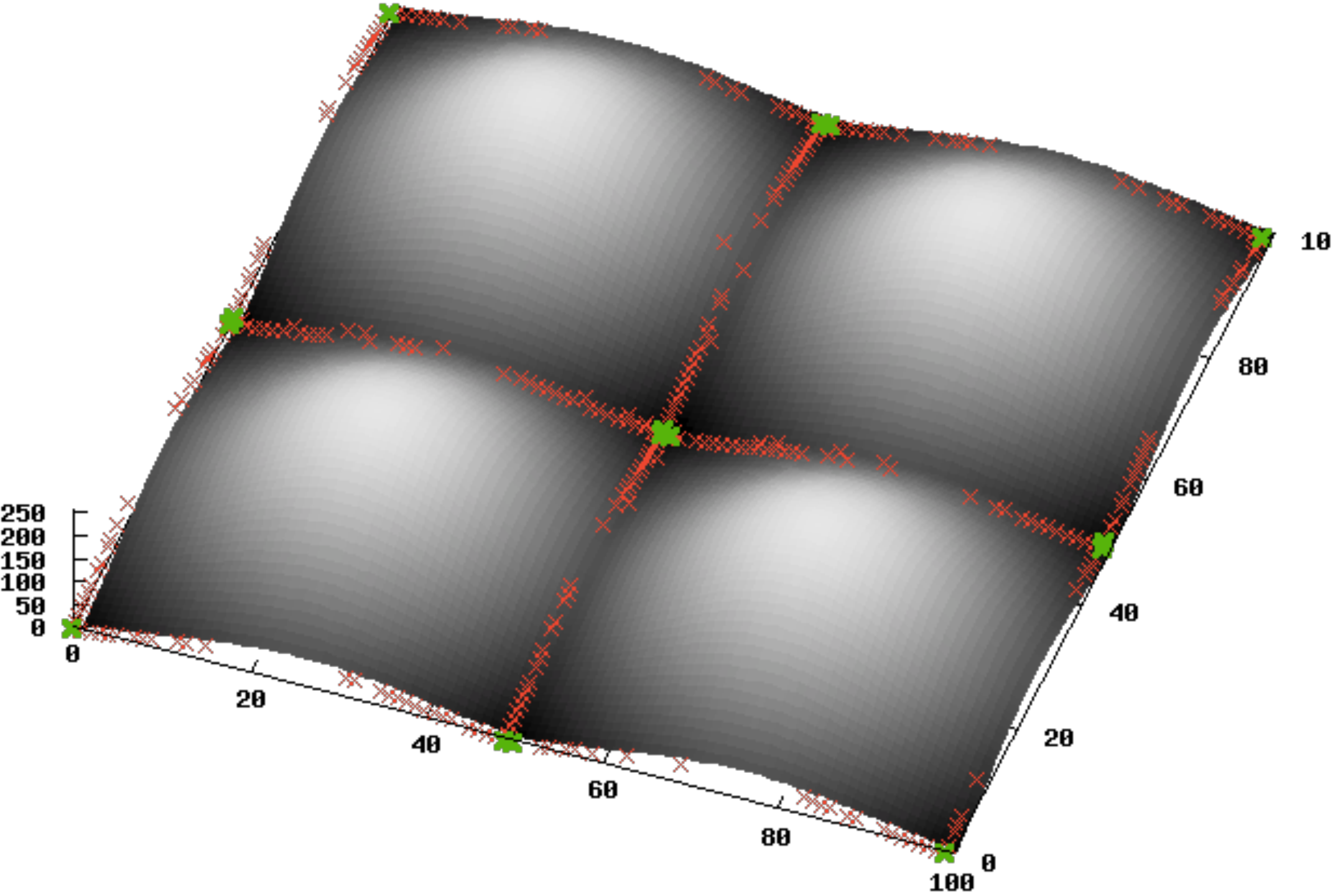}
\caption{The final position of 1000 artificial crawlers, (a) by using the rule of movement $max$ and (b) by using the rule of movement $min$. Green marks stand for live artificial crawlers while red marks represent dead artificial crawlers.}
\label{fig:maxmin}
\end{figure*}

In the first step, the artificial crawlers move to higher intensities as the original method.
Thus, artificial crawlers $A^i_{T} = \{ p_{T}^i, e_T^i \}$ are obtained after the evolution process converges, where $T$ is the number of step needed to the system converges.
Throughout the paper, the artificial crawlers which move to higher intensities will be referred to as $U_{T}^i = \{p_{T}^i, e_T^i \}$ and this rule of movement will be referred to as $max$.
Figure \ref{fig:maxmin} shows an example of 1000 artificial crawlers using the rule of movement $max$. Although the image in Figure  \ref{fig:maxmin}(a) is an elaborate example we present how the agents find the maxima accordingly the $max$ rule. 
The green marks stand for the final position (convergence) of the live artificial crawler while the red ones represent the final position of the dead artificial crawlers.
As we can see, the live artificial crawlers can achieve the highest intensities.
The energy of these artificial crawlers implicitly stores all the information along the steps and this energy is important to represent the surface that the artificial crawler is emerged.
As important as the live artificial crawler's, the dead ones aggregate information from the surface of the environment.

In Figure \ref{fig:maxmin}, we can observe that the original method only describes the peaks of a texture image.
In addition, we propose to move artificial crawlers toward lower intensities.
In this approach, artificial crawlers $Q_{t}^i = \{ p_{T}^i, e_T^i \}$ are randomly placed in the image with initial energy $e_{init}$.
Then, the evolution process is modified so that the next step of an artificial crawler is to move towards the lower intensity (Equation \ref{eq:min}).
This rule of movement will be referred to throughout the paper as $min$.

\begin{equation}
f(p_t^i) = \left\{ \begin{array}{lll}
         p_t^i, & \textrm{if } I(p_t^i) \le I(p) \, \forall p \in \eta(p_t^i) \\
         p, & \textrm{if } I(p) < I(p_t^i), I(p) < I(q) \, \forall p, q \in \eta(p_t^i), \\
         & p \neq q \\
         p, & \textrm{if } I(p) < I(p_t^i), I(p) \leq  I(q) \, \forall p, q \in \eta(p_t^i), \\
         & p \neq q, p \textrm{ was visited}
         \end{array} \right. 
         \label{eq:min}
\end{equation}

An example of the artificial crawlers using the rule of movement $min$ can be seen in Figure \ref{fig:maxmin} (b).
Again, green marks represent the final position of live artificial crawlers while red marks represent the dead artificial crawlers.
These artificial crawlers complement the artificial crawlers that use the rule of movement $max$, aggregating more information about the surface.


In the end of this step, we have two populations of $N$ artificial crawlers $U_{T}^i = \{p_{T}^i, e_T^i \}$ and $Q_{T}^i = \{p_{T}^i, e_T^i \}$ which correspond to the artificial crawlers using rules of movement $max$ and $min$, respectively.

\subsection{Fractal Dimension of Artificial Crawler}

\begin{figure*}[!htbp]
\centering
\subfigure[Artificial crawlers mapped onto a 3D space by converting the final position and the energy into a point in the surface.]{\includegraphics[width=0.80\textwidth]{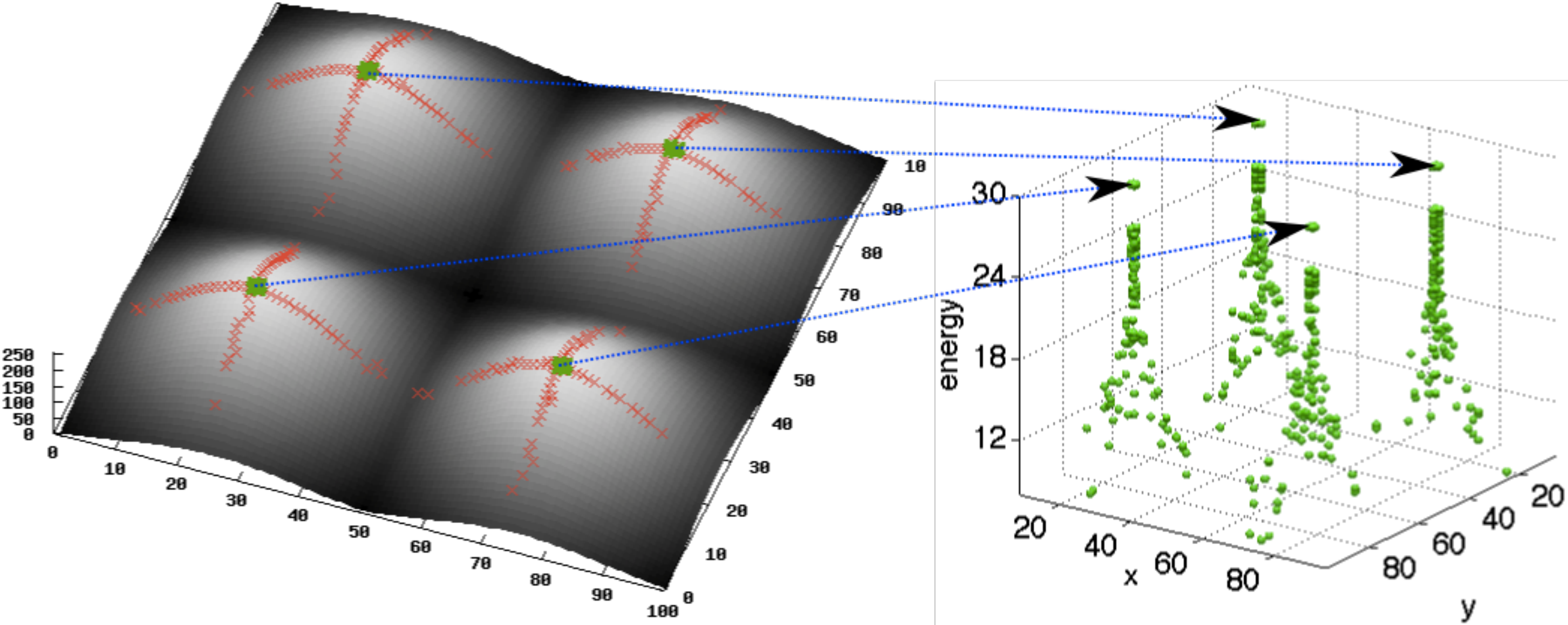}} \\
\subfigure[$r=2$]{\includegraphics[width=0.35\textwidth]{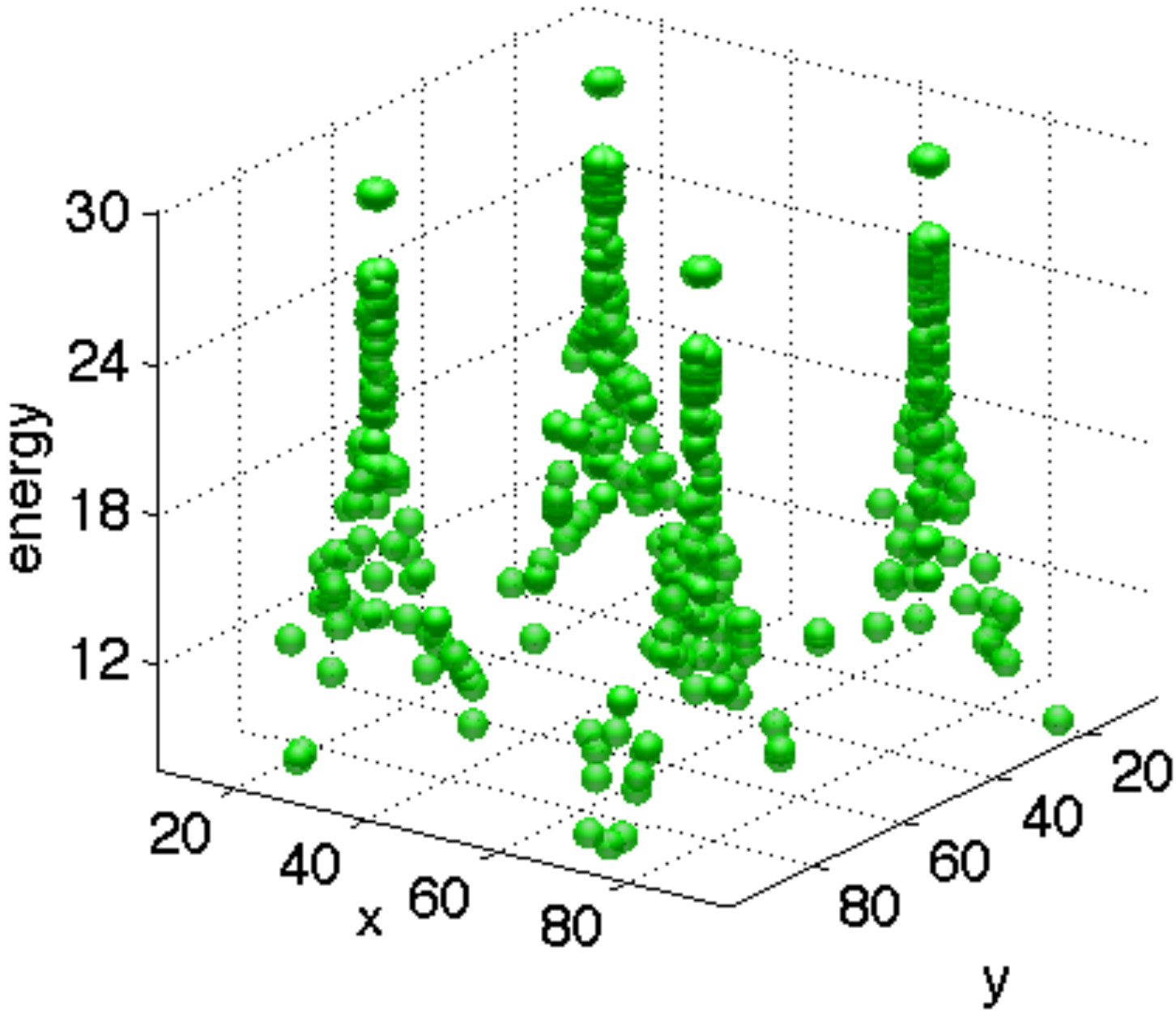}}
\subfigure[$r=3$]{\includegraphics[width=0.35\textwidth]{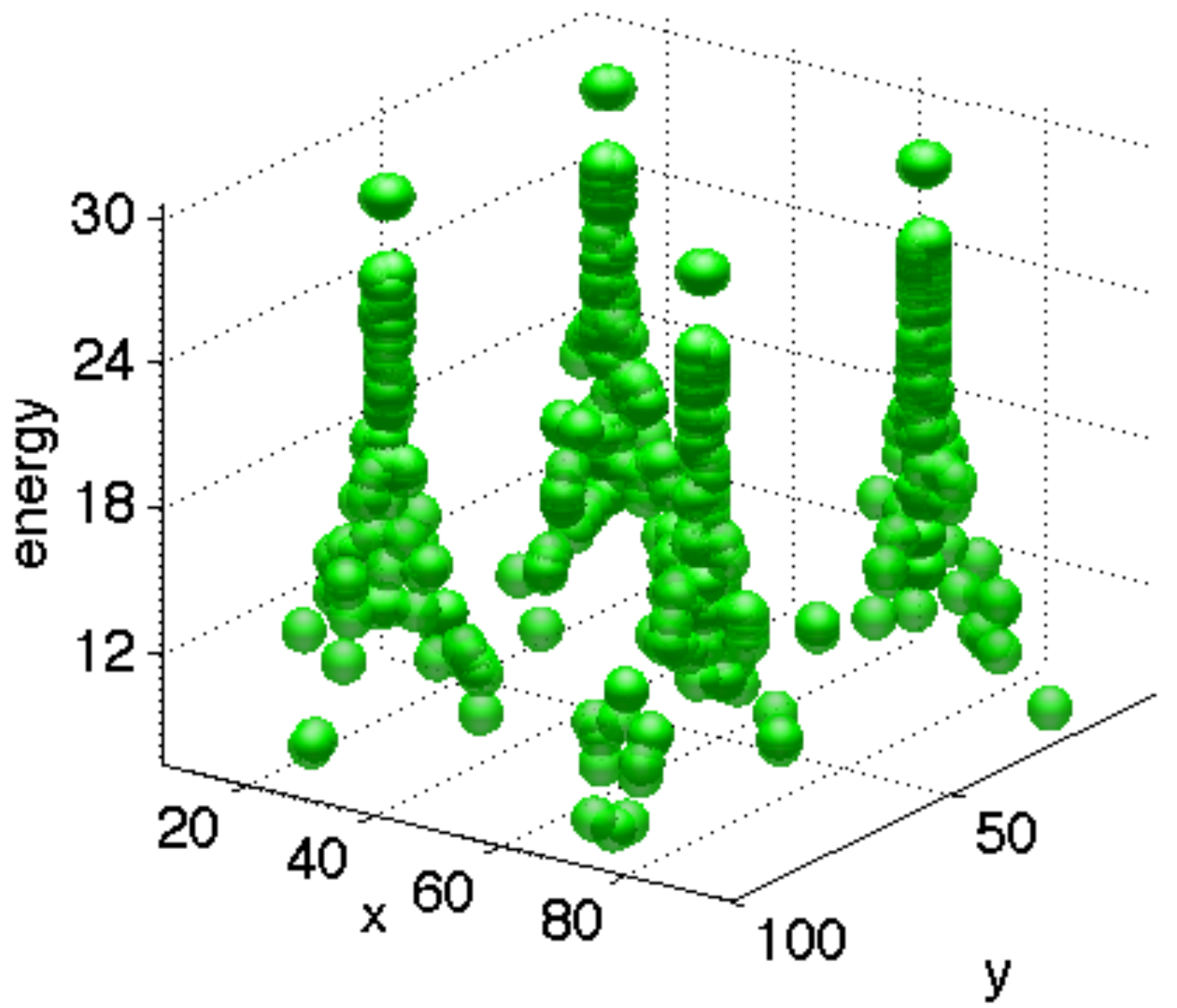}}
\caption{An illustration of the dilation process for the fractal dimension estimation of artificial crawlers. The final position of the artificial crawlers was obtained using the rule of movement $max$ and maximum energy $e_{max} = 30$.}
\label{fig:fractal_dilation}
\end{figure*}

In this section, we describe how to quantify the population of artificial crawlers using the fractal dimension theory.
To estimate the fractal dimension using the Boulingand-Minkowski method, the population of artificial crawlers can be easily mapped onto a surface $S \in \Re^3$, by converting the position $p_T^i = \{ x_i, y_i\}$ and the energy $e_T^i$ of each artificial crawler into a 3D point $s_i =(x_i,y_i,e_T^i)$.
The energy is important because it contain all the information of the artificial crawler steps.
This mapping can be seen in Figure \ref{fig:fractal_dilation} (a).
We should note that the $Z$ axis is the energy of the artificial crawler.

The Boulingand-Minkowski method estimates the fractal dimension based on the size of the influence area $|S(r)|$ created by the dilation of $S$ by a radius $r$.
Thus varying the radius $r$, the fractal dimension of surface $S$ is given by:

\begin{equation}
 D = 3 - \lim_{r \to 0} \frac{log \, V(r)}{log \, r}
 \label{eq:dimFractal2}
\end{equation}
where $V(r)$ is the influence volume obtained through the dilation process of each point of $S$ using a sphere of radius $r$:
\begin{equation}
 V(r) = |\{ s' \in \Re^3 \mid \exists s \in S: |s-s'| \leq r \}|
\end{equation}

The dilation process is illustrated in Figure \ref{fig:fractal_dilation}.
A group of artificial crawlers are mapped onto a 3D space, shown in Figure \ref{fig:fractal_dilation} (a).
Each point of the 3D space is dilated by a sphere of radius $r$ (Figure \ref{fig:fractal_dilation} (b) and (c)).
As the value of radius $r$ is increased, more collisions are observed among the dilated spheres.
These collisions disturb the total influence volume $V(r)$, which is directly linked to the roughness of the surface.

From the linear regression of the plot of $\log r \times \log V(r)$, the Boulingand-Minkowski fractal dimension is computed by:
\begin{equation}
D = 3 - \alpha
 \label{eq:dimFractal3}
\end{equation}
where $\alpha$ is the slope of the estimated line.

%

\subsection{Feature Vector}
\label{sec:feature-vector}
Although the fractal dimension provides a robust mathematical model, it describes each object by only one real value $D$ $-$ the fractal dimension.
Objects with distinct shapes can have the same fractal dimension, for instance, the very well know fractals :Peano curve, Dragon curve, Julia set and the boundary of the Mandelbrot set have the same Hausdorff  dimension equals to 2. To overcome this characteristic the concept of multi-scale fractal dimension \cite{odemirIS2008} and the fractal descriptors \cite{BackesCB12,FlorindoBCB12} was developed. In this way, the fractal dimension of the object is considered in different scales. It provides a rich shape descriptor that can be successful to discriminate shape and patterns \cite{odemirIS2008}.


In \cite{odemirIS2008} it was demonstrate 

In order to improve the discrimination power of our method, we use the entire curve $V(r)$ instead of using only the fractal dimension:

\begin{equation}
\varphi_{\tau} = [V(1), \dots, V(r_{m}) ]
\label{eq:vetor}
\end{equation}
where $\tau$ is the rule of movement used by the artificial crawler and $r_{m}$ is the maximum radius.

Considering that we have two rules of movement, the final feature vector is composed by the concatenation of $\varphi_{max}$ and $\varphi_{min}$ according to Equation \ref{eq:finalvector}. 
The feature vectors $\varphi_{max}$ and $\varphi_{min}$ are obtained by using the fractal dimension estimation of acrawlers $U_T^i$ and $Q_T^i$ after the stabilization, respectively.
\begin{equation}
\varphi = [\varphi_{max}, \varphi_{min} ]
\label{eq:finalvector}
\end{equation}

\begin{figure*}[!htbp]
\centering
\subfigure[Classes of texture D16, D18, D49 and D93.]{\includegraphics[width=0.7\textwidth]{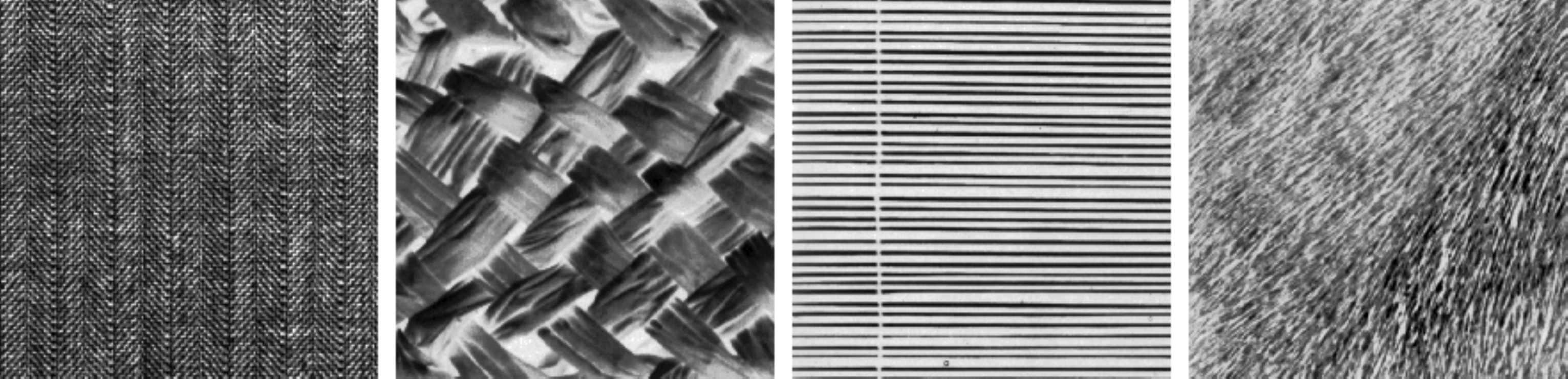}}
\subfigure[$\varphi_{max}$ for classes D16 and D18.]{\includegraphics[width=0.38\textwidth]{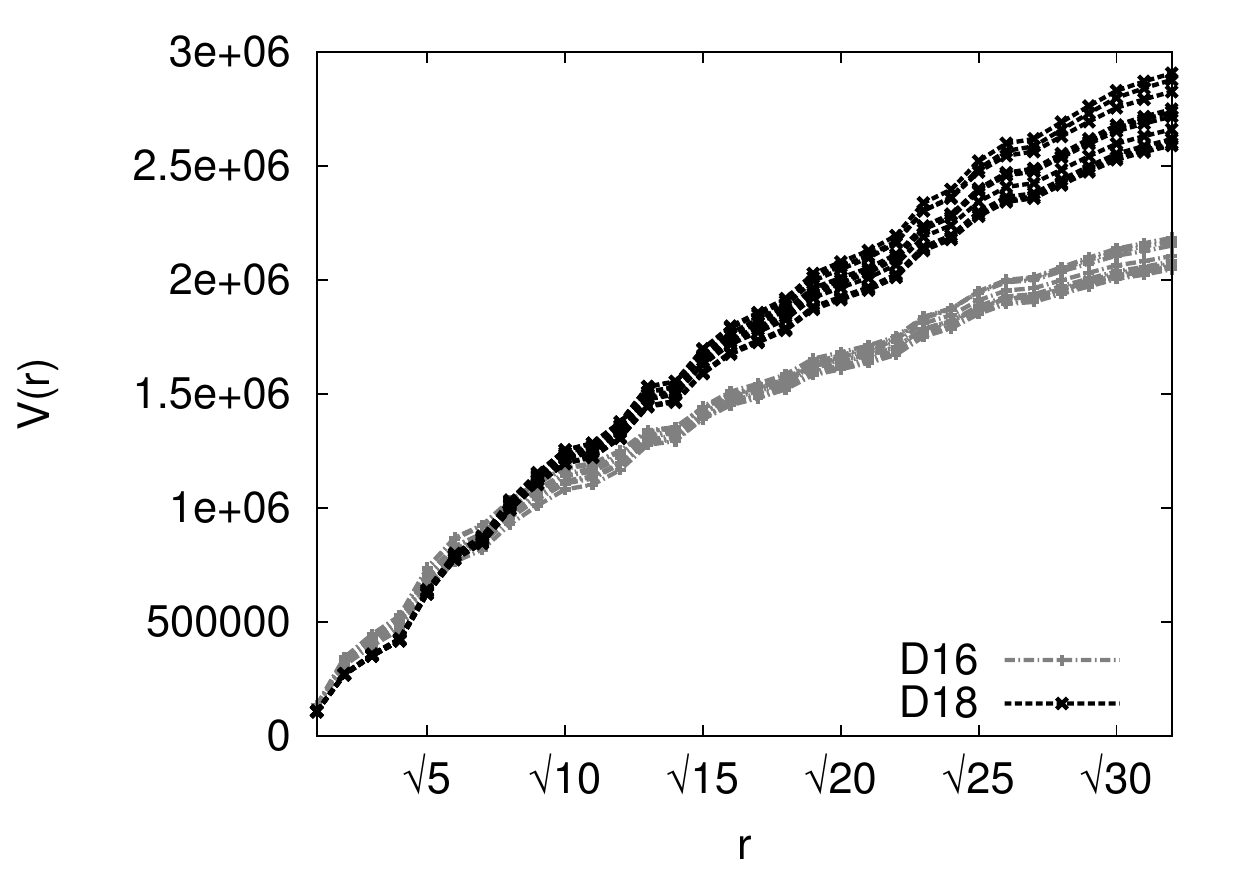}}
\subfigure[$\varphi_{min}$ for classes D16 and D18.]{\includegraphics[width=0.38\textwidth]{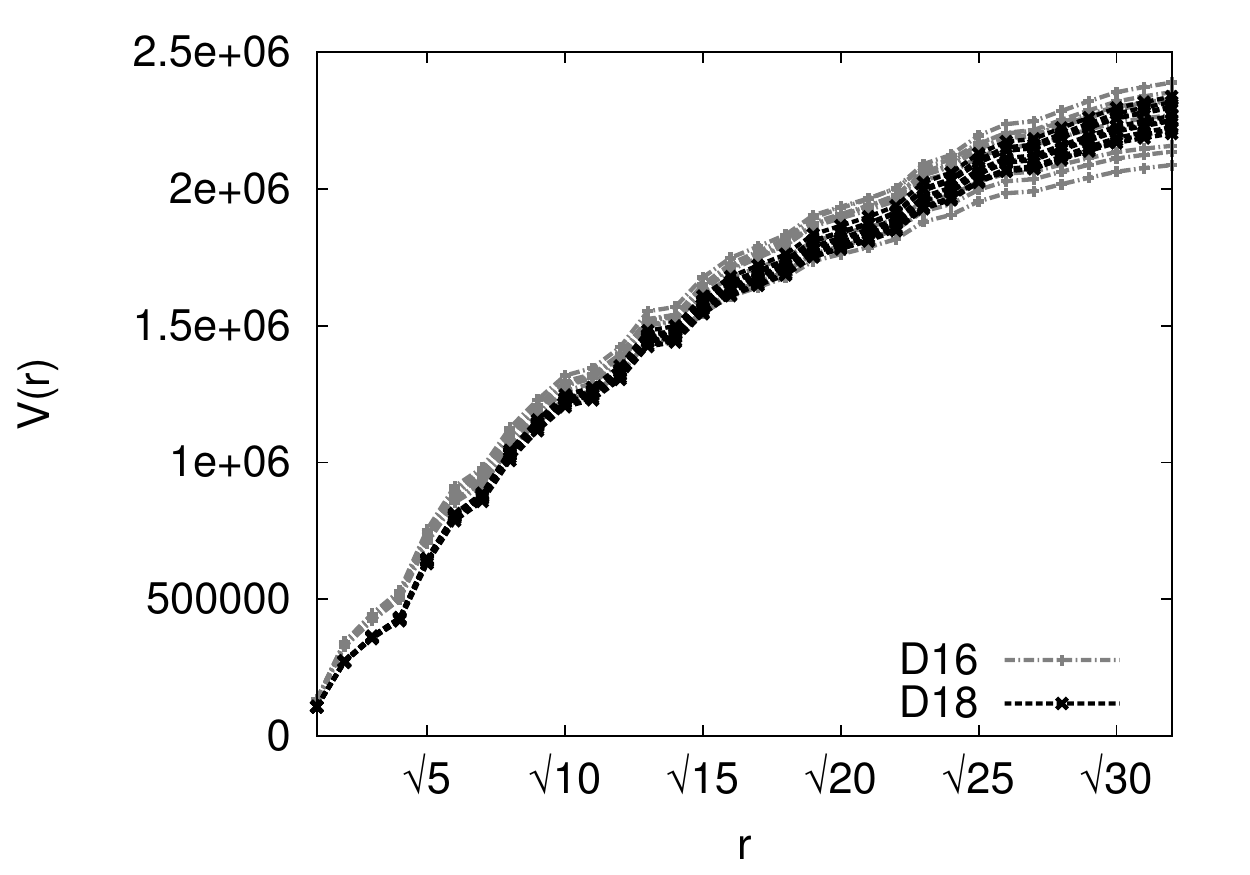}}
\subfigure[$\varphi_{max}$ for classes D49 and D93.]{\includegraphics[width=0.38\textwidth]{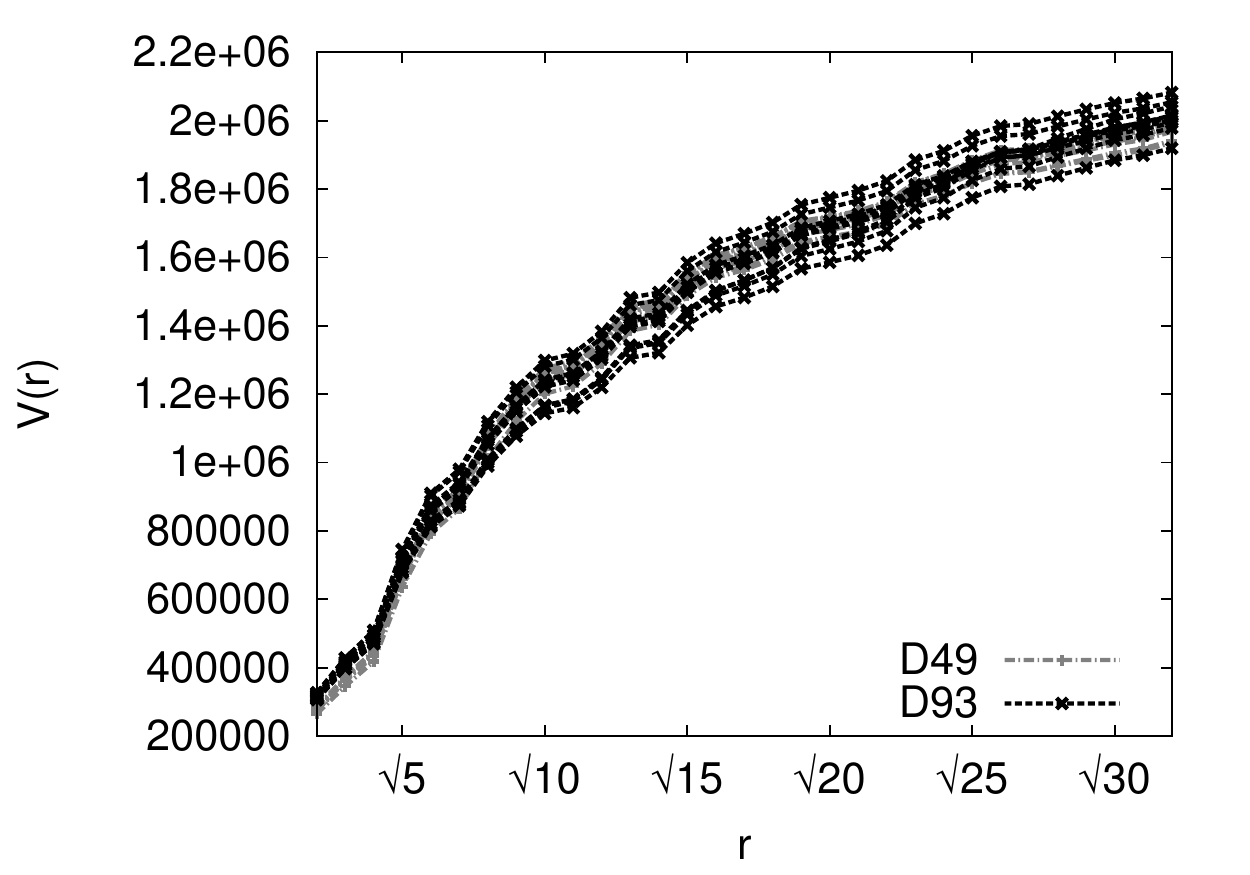}}
\subfigure[$\varphi_{min}$ for classes D49 and D93.]{\includegraphics[width=0.38\textwidth]{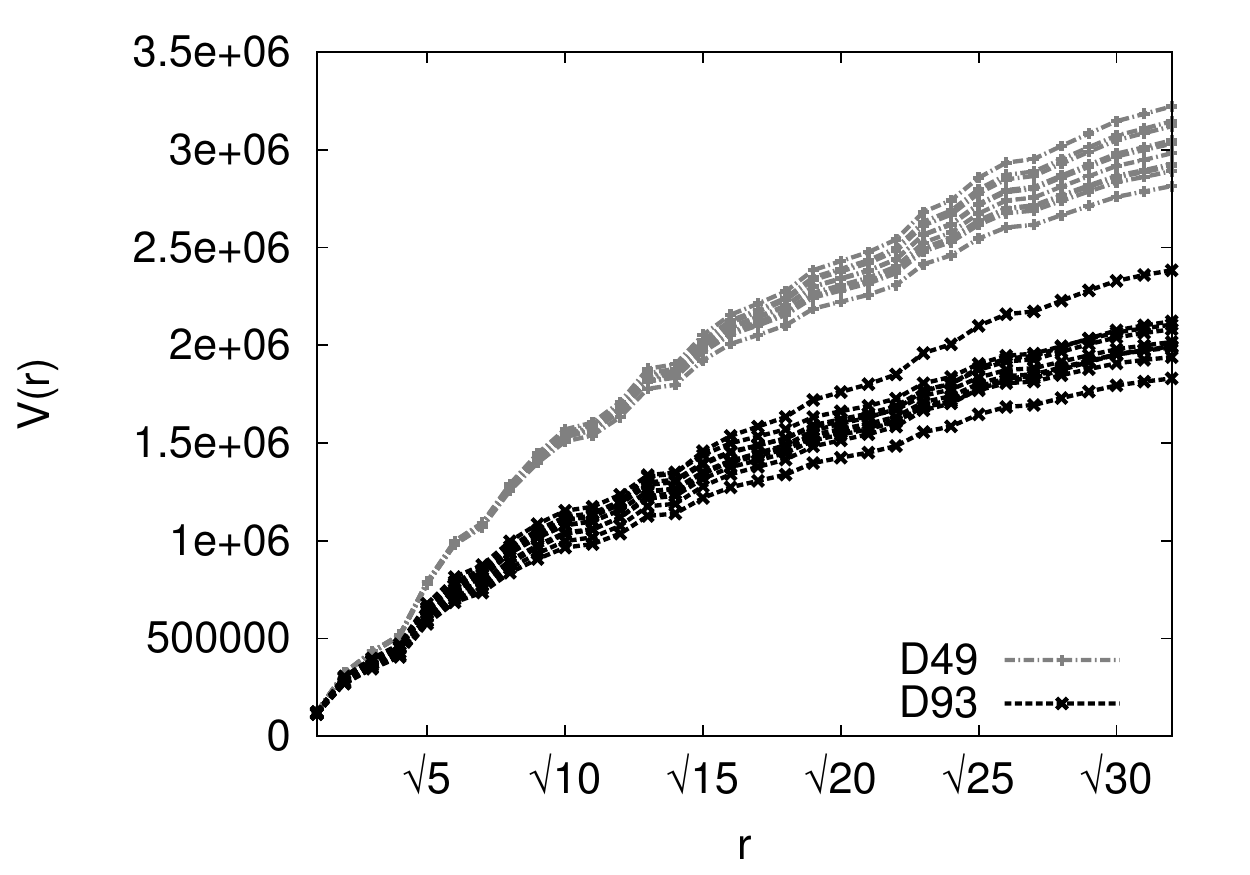}}
\caption{An example of feature vectors using the rules of movement $min$ and $max$. The classes of texture are only discriminated if both rules are used.}
\label{fig:minmax1}
\end{figure*}

The importance of using both rules is corroborated in Figure \ref{fig:minmax1}.
Figures \ref{fig:minmax1} (b) and (d) show the feature vectors by using $\varphi_{max}$ only, and Figures \ref{fig:minmax1} (c) and (e) show the feature vectors by using $\varphi_{min}$ only.
An example of those feature vectors are obtained for four different image classes, as shown in Figure \ref{fig:minmax1} (a). For clarify, each class contains 10 samples. The classes D16 and D18 are discriminated using the rule of movement $max$ (Figure \ref{fig:minmax1} (b)), while the rule of movement $min$ is not able to discriminate those two classes accordingly (Figure \ref{fig:minmax1} (c)).
On the other hand, the classes D49 and D93 are only discriminated if the rule of movement $min$ is used (Figure \ref{fig:minmax1} (e)).
These plots corroborate the importance of using both rules of movement for texture recognition.

\subsection{Computational Complexity}
In the proposed method, $N$ artificial crawlers are performed in the image of size $W \times W$ pixels.
The swarm system converges after $M$ steps, which leads to a computational complexity of $O(NM)$.
After stabilization, we propose to quantify the swarm system by means of the fractal dimension.
To calculate the dilation process, the Euclidean distance transform \cite{fabbri2008} is a powerful and efficient tool.
This transform calculates the distance between each point of the 3D space and the surface.
Several authors \cite{saito1994,meijster2000,fabbri2008} proposed algorithms for computing Euclidean distance transform in linear time.
The time complexity is linear in the number of points of the 3D space, which is $O(W \times W \times e_{max})$ $-$ $W \times W$ is the size of the image and $e_{max}$ is the maximum energy of the agents.
Usually, the maximum energy $e_{max}$ is a small number (e.g. in this work the maximum energy is $20$).
Thus, we can ignore $e_{max}$ in the complexity, since $W \gg 20$ in image applications.
Finally the computational complexity of the proposed method is stated as $O(NM + W^2)$.

\begin{figure}[!htbp]
   \begin{center}
     \includegraphics[width=0.7\columnwidth]{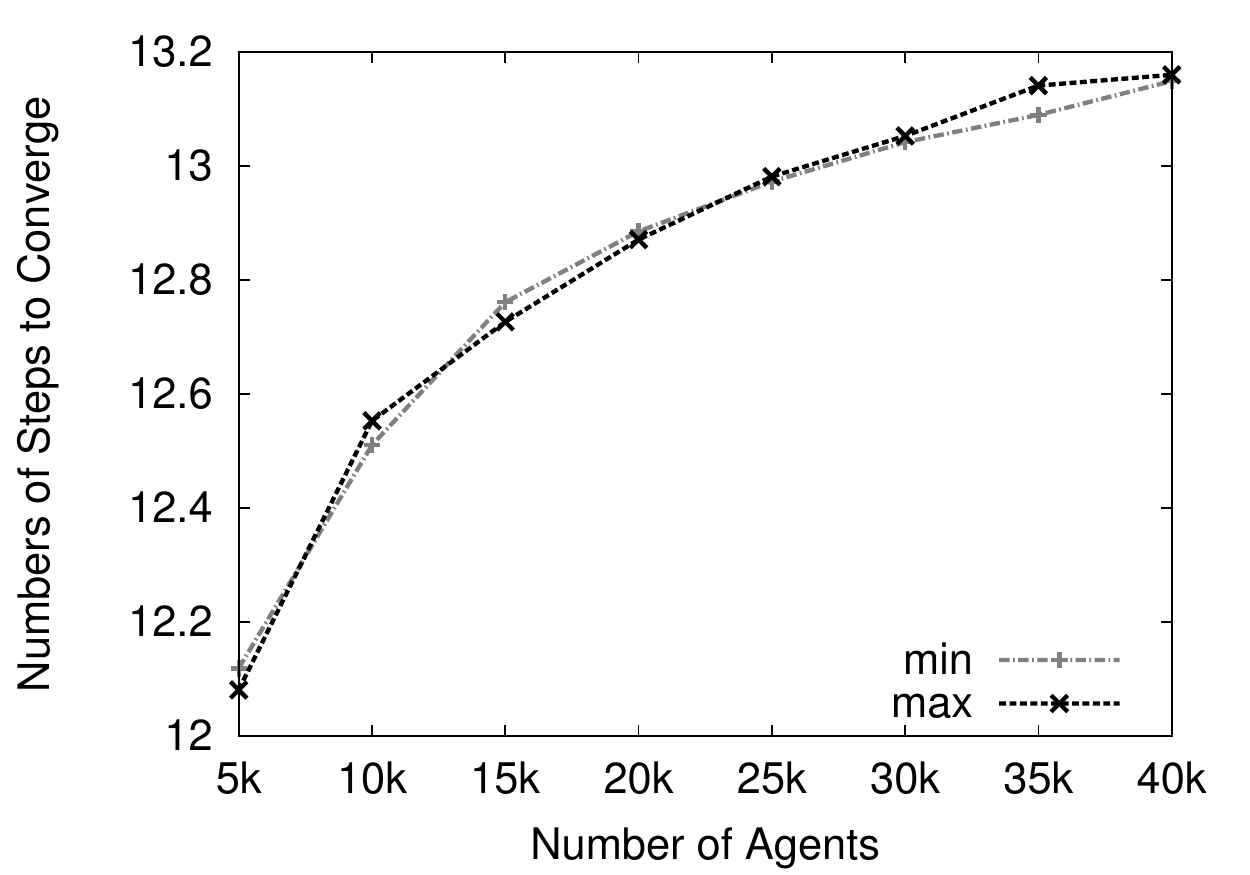}
     \caption{Average number of steps to converge using the rules of movement $min$ and $max$. The average number of steps was averaged over 400 images.}
     \label{fig:stepstoconverge}
   \end{center}
\end{figure}

Let us discuss the best, worst and average case based on the number of steps of the swarm system.
The best case considers that the swarm system converges in one step ($M=1$).
Thus, the computation complexity is $O(N + W^2)$.
In the worst case, the swarm system takes more than $N$ steps, however it is stopped in $M=N$ steps without the stabilization.
The worst case leads to a complexity of $O(N^2 + W^2)$.
It is important to emphasize that the worst case rarely occurs, requiring a specific configuration of the texture image and even a random image does not produce this special case.
In order to analyze the average case, we plot in Figure \ref{fig:stepstoconverge} the average number of steps needed to converge over 400 images.
We can see that the two rules of movement $min$ and $max$ present similar behavior.
Also, the number of agents does not influence the number of steps to converge (e.g. the difference of the average number os steps for $N = 5k$ and $N=40k$ is only $1.03$ steps).
Given that $M \sim 13$ for $N=50k$, the average case leads to a complexity which is very close to the best case, $O(N + W^2)$, and it is a good complexity in comparison to the complexities of Gabor filters $O(W^2 \log W)$ and co-occurrence matrices O($W^2$).

\section{Experimental Results}
In order to evaluate the proposed method, experiments were carried out on image datasets of high variability. We first describe such datasets, the experiments to evaluate the parameters of our proposed method, and then the comparative results with the state-of-the-art methods.

\begin{figure}[!htbp]
   \begin{center}
     \includegraphics[width=0.8\columnwidth]{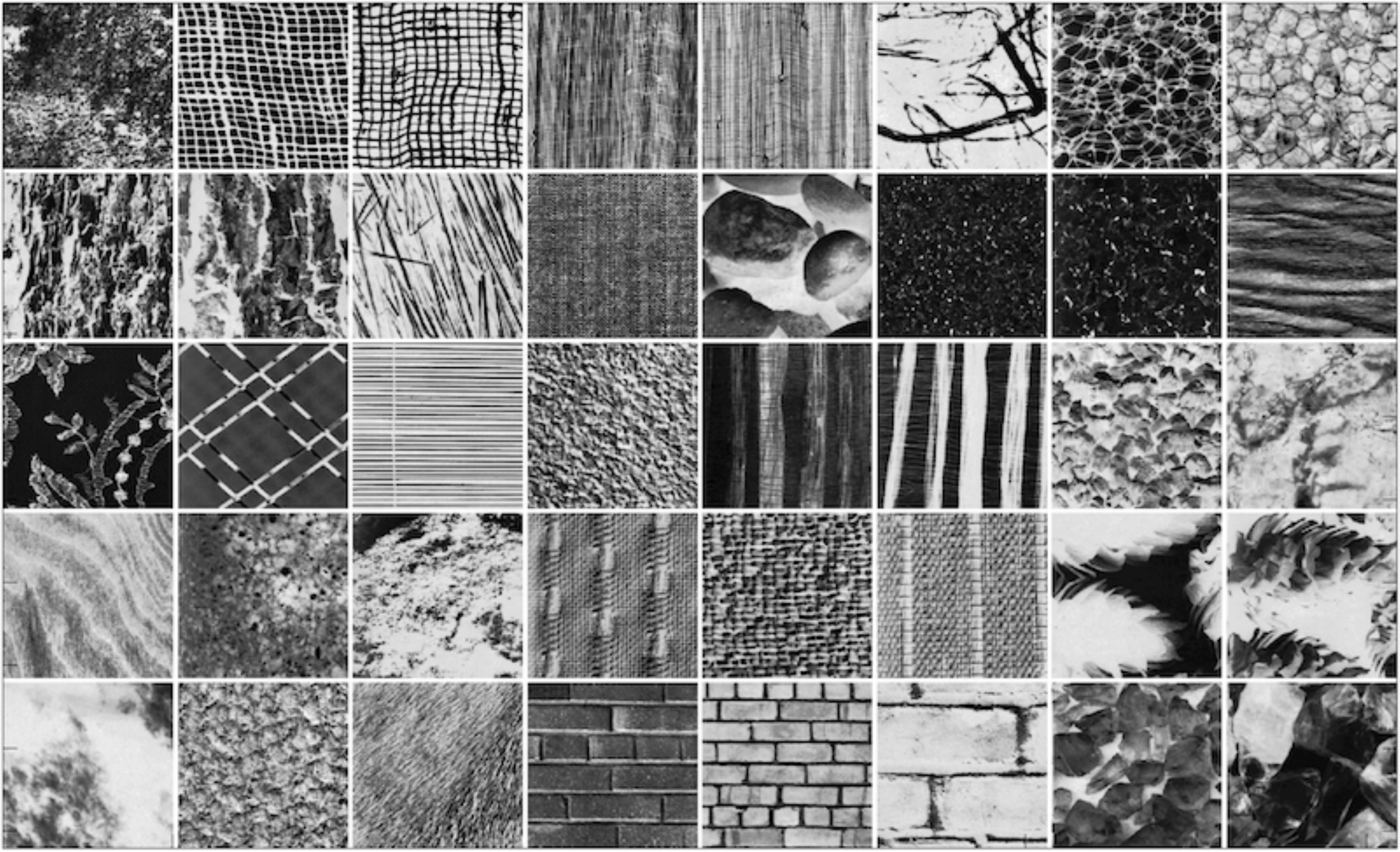}%
     \caption{An example of the classes from the Brodatz dataset. Each class has 10 samples of $200 \times 200$ pixels and 256 gray levels.}
     \label{fig:brodatz}
   \end{center}
\end{figure}

We performed experiments on the two most used image datasets of texture: Brodatz and Vistex.
The Brodatz album \cite{Brodatz1966} is the most known benchmark for evaluating texture methods. Each class is composed by one image divided into ten non-overlapped samples. The samples have $200\times200$ pixels with 256 gray levels. In this work, a total of 40 classes with 10 samples per class were used. One example of each class is shown in Figure \ref{fig:brodatz}. 

\begin{figure}[!htbp]
   \begin{center}
     \includegraphics[width=0.8\columnwidth]{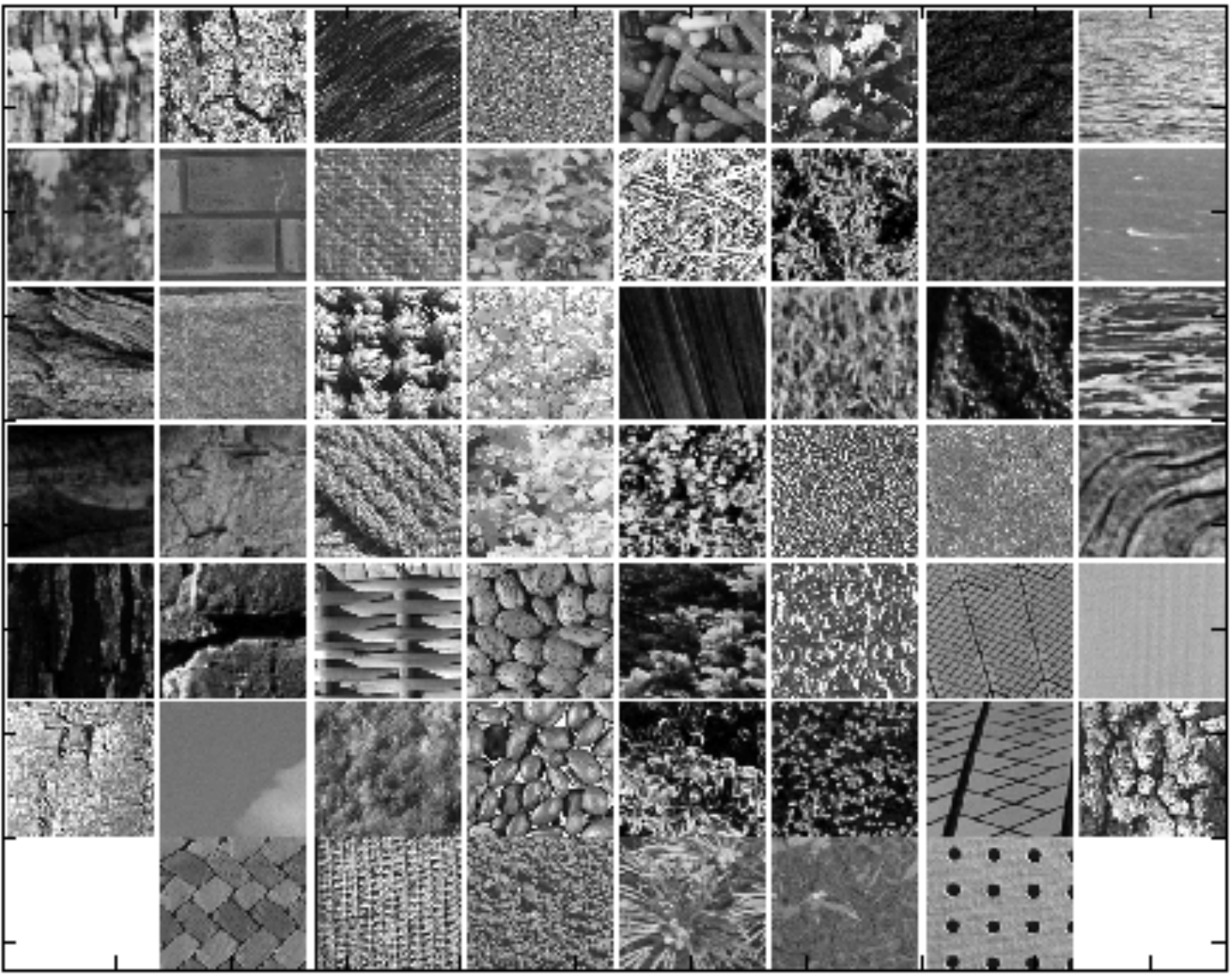}
     \caption{An example of the classes from the Vistex dataset. Each class has 16 samples of $128 \times 128$ pixels and 256 gray levels.}
     \label{fig:vistex}
   \end{center}
\end{figure}

The Vision Texture $-$ Vistex \cite{SinghICAPR2001} provides real-world textures under challenging conditions (e.g. lighting and perspective).
A total of 54 classes are available, each class containing 16 samples.
The samples have $128 \times 128$ pixels with 256 gray levels.
Figure \ref{fig:vistex} presents one example of each class. 

In our experiments, Linear Discriminant Analysis (LDA) \cite{timm2002applied,Fukunaga1990} in a 10-fold cross-validation strategy was adopted in the task of classification.
The LDA method estimates a linear subspace in which the projection of the vectors presents larger variance inter-classes than the variance intra-classes.
The 10-fold cross-validation strategy divides randomly the samples into 10 folds. Each fold is used to test the classifier while the other nine folds are used to train the classifier. This process is repeated 10 times with each fold used once as testing data. To produce a single statistic, the results of the 10 processes are averaged.

The features used in this paper, and the parameter evaluation are in the next section.

\subsection{Parameter Evaluation}
In this section, we evaluate the three main parameters of our method: number of artificial crawlers $N$, maximum energy $e_{max}$ and maximum radius $r_{m}$ of the fractal dimension.
The other parameters were set according to \cite{ZhangIJPRAI2005}, since their possible values do not affect the final success rate.
Each artificial crawler is born with initial energy $e_{init} = 10$, the survival threshold $e_{min} = 1$ and the absorption rate $\lambda = 0.01$.

The success rates for the different number of artificial crawlers are shown in Figure \ref{fig:n} for both Brodatz and Vistex datasets.
The number of artificial crawlers placed on the pixels were initially set to $5k$ with a coverage rate of 5\%, varying from $5k$ to $40k$ for the Brodatz dataset and varying from $5k$ to $15k$ for the Vistex dataset due to the size of the samples ($128 \times 128$ pixels).
We can observe that the highest success rate was obtained for $N=30k$ and $N=15k$ for the Brodatz and Vistex, respectively.
Further, it was found that the combination of rules $min$ and $max$ significantly improve the success rate for all number of artificial crawlers in both datasets.
Also, the rule of going to the minimum intensity provides similar results to the original rule $-$ $max$.
These results suggest that the valleys and peaks are important to obtain a robust texture analysis.

\begin{figure*}[!htbp]
\centering
\subfigure[Brodatz]{\includegraphics[width=0.48\textwidth]{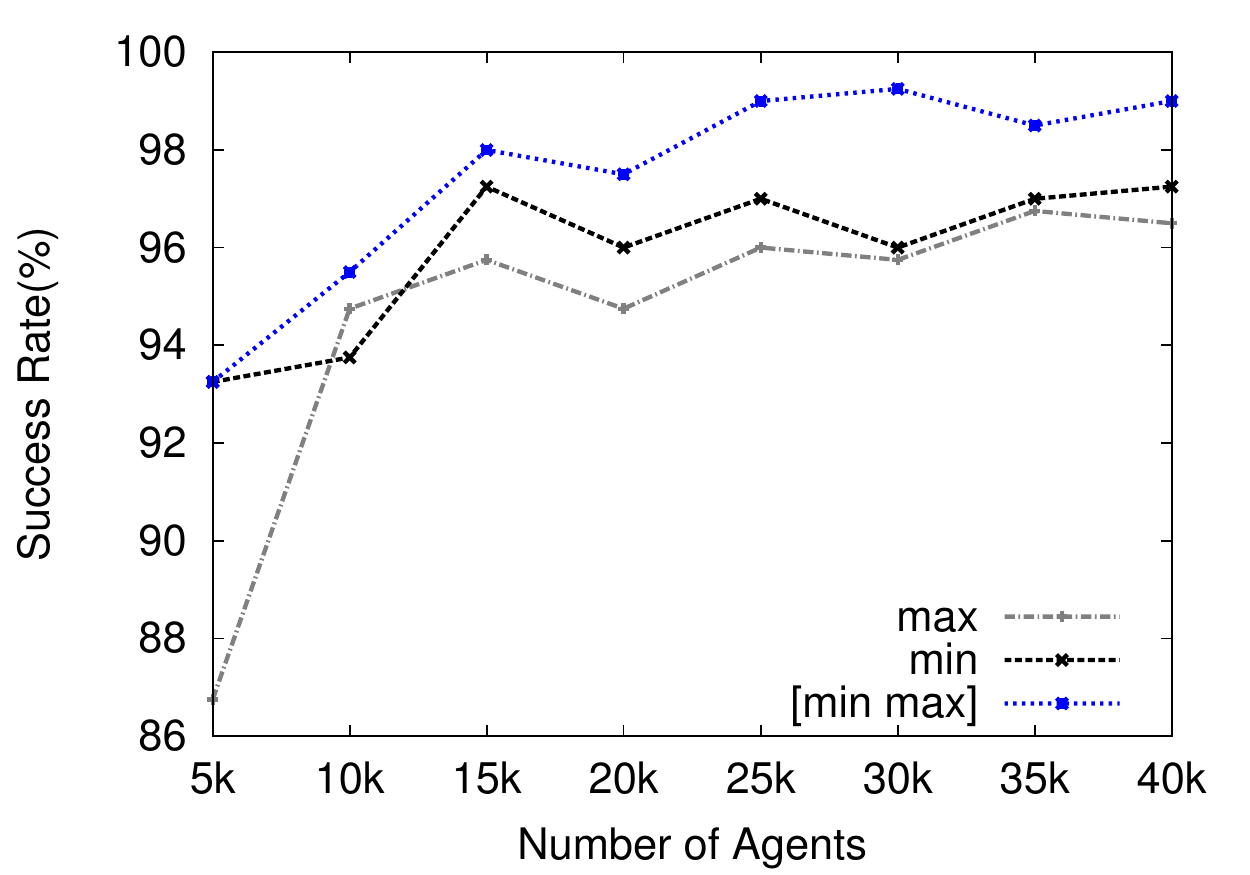}}
\subfigure[Vistex.]{\includegraphics[width=0.48\textwidth]{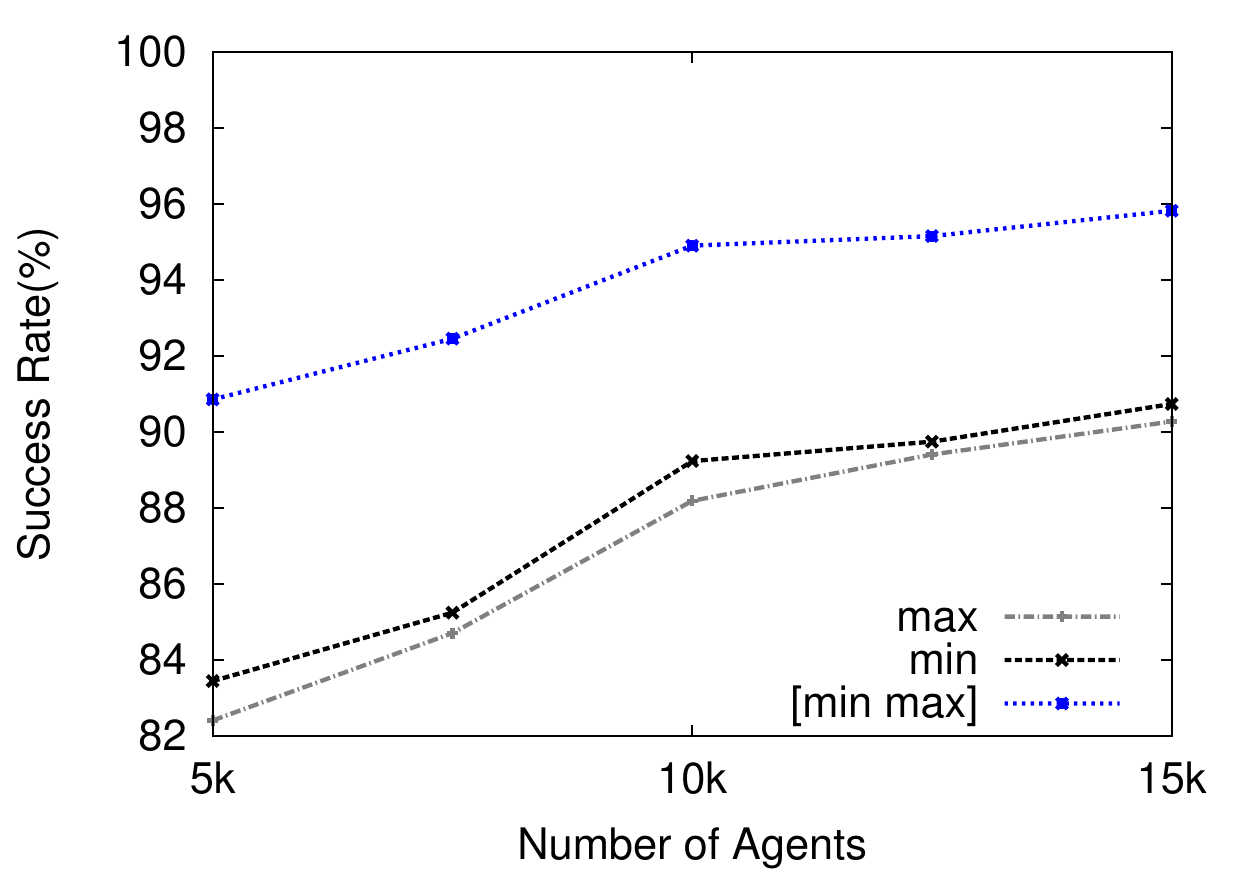}}\\
\subfigure[Brodatz]{\includegraphics[width=0.48\textwidth]{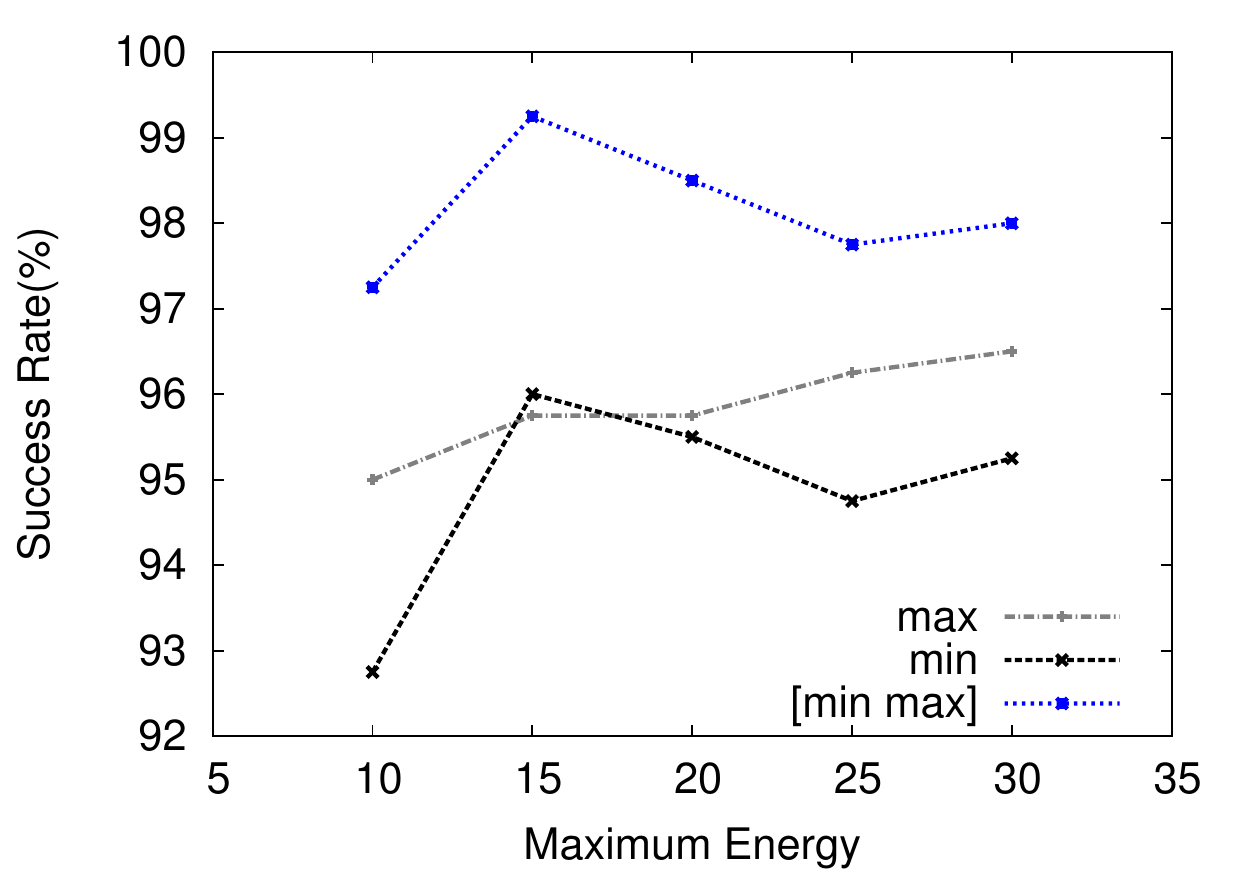}}
\subfigure[Vistex.]{\includegraphics[width=0.48\textwidth]{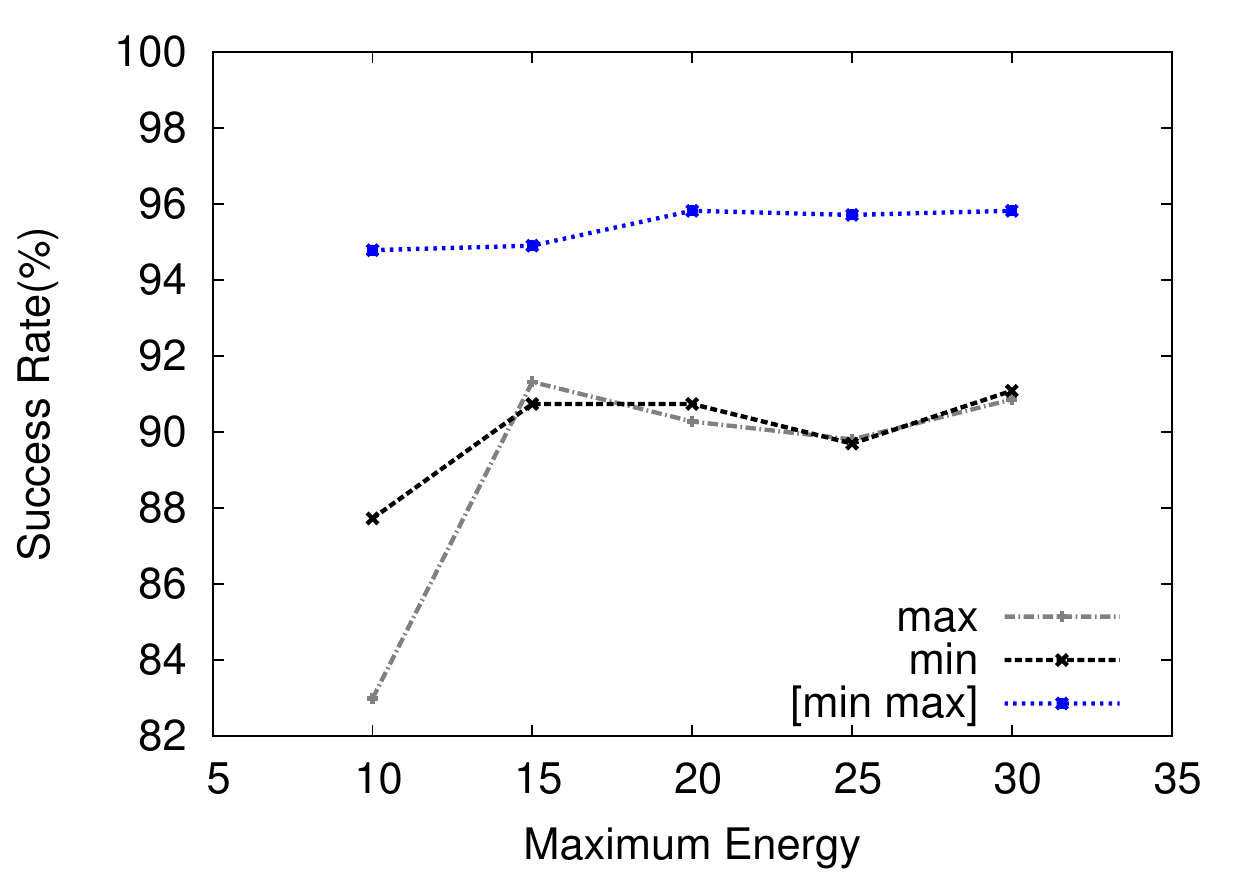}}\\
\subfigure[Brodatz]{\includegraphics[width=0.48\textwidth]{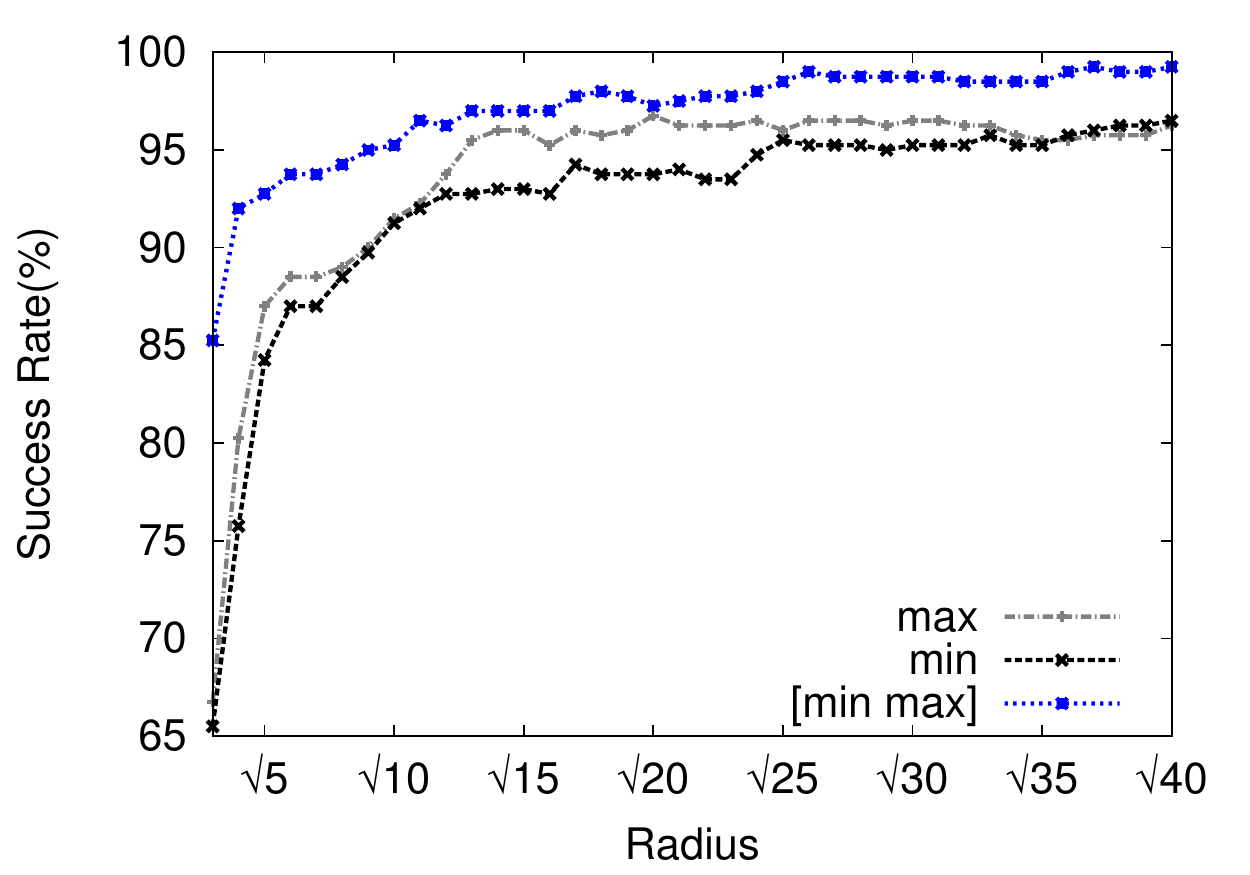}}
\subfigure[Vistex.]{\includegraphics[width=0.48\textwidth]{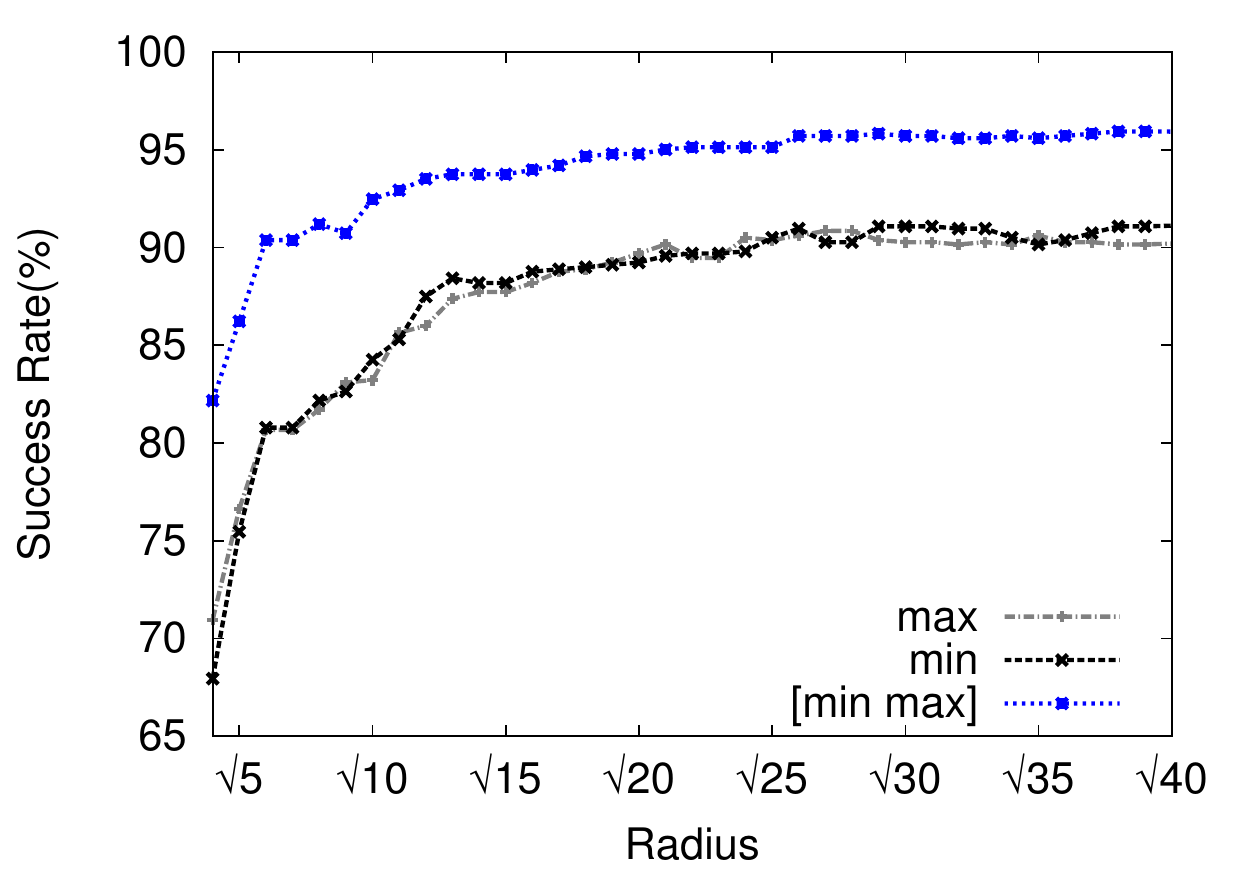}}
\caption{Plots for different The plot for evaluating the number of artificial crawlers in the Brodatz and Vistex datasets.}
\label{fig:n_energy_radius}
\end{figure*}

%
%

The maximum energy of the artificial crawlers is evaluated in the plot of the Figure \ref{fig:n_energy_radius}.
Figure \ref{fig:n_energy_radius} (c) presents the results for the Brodatz dataset while Figure \ref{fig:n_energy_radius} (d) shows the results for the Vistex dataset.
The maximum energy parameter was evaluated by the fact that it limits the artificial crawler energy and, consequently, can limit the fractal dimension space.
However, the experimental results show that different values of maximum energy do not influence the success rate considerably.
The highest success rate was obtained for $e_{max} = 15$ using the Brodatz dataset and for $e_{max} = 20$ using the Vistex dataset.
It can be noted that the same behavior for the rules of movement was obtained here, with the combination of rules providing the highest success rates.

In the plot of Figure \ref{fig:n_energy_radius}, the maximum radius of the fractal dimension estimation is evaluated.
As expected, the success rate increases as the radius increases and stabilizes after a certain radius.
The maximum radius $r_{m} = \sqrt{37}$ provided the highest success rate of $99.25\%$ for the Brodatz dataset.
For the Vistex dataset, a success rate of $95.95\%$ was obtained by the maximum radius $r_{m} = \sqrt{38}$.
As the previous results, the combination of rules of movement provides the highest success rates.
Also, the rule $min$ provides similar results compared to the rule $max$.
Using these plots, we can set the parameters of the proposed method for the Brodatz dataset to $N=30k$, $e_{max} = 15$, $r_m = \sqrt{37}$.
For the Vistex dataset, the parameters are $N=15k, e_{max} = 20$ and $r_m = \sqrt{38}$, which are close to the best parameters for the Brodatz dataset.
For other datasets, we recommend using a number of agents $N$ between $60\%$ and $95\%$ of the number of pixels, $10 \leq e_{max} \leq 25$ and $ \sqrt{30} \leq r_m \leq \sqrt{40}$.

\subsection{Comparison with other Methods}
The proposed method, which is enriched by the fractal dimension estimation of artificial crawler, is compared to traditional texture methods, namely Fourier descriptors \cite{azencottPAMI1997}, co-occurrence matrices \cite{ChristophPR2004,haralickTSMC1973}, Gabor filter \cite{Bianconi2007,Jain1991,gaborJIEE1946}, local binary pattern \cite{Ojala2002}, and multi fractal spectrum \cite{Xu2009}.
Moreover, the texture method using the artificial crawlers proposed in \cite{ZhangIJPRAI2005} was also used in this comparison.
We considered the traditional implementation of each method and its parameter configuration as described below, which yields the best result.

\textit{Fourier descriptors}: these descriptors are obtained from the Fourier transform of the texture image. Each descriptor is the sum of the spectrum values within a radius from the center.  The best results were obtained by radius with increment by one. Thus, for an image of $200 \times 200$ pixels, 99 descriptors are obtained. More information about the Fourier descriptors can be found in \cite{azencottPAMI1997}.

\textit{Co-occurrence matrices}: they are computed by the joint probability distribution between pairs of pixels at a given distance and direction. In these experiments, we consider the distances from 1 to 5 pixels, and the angles $0^o$, $45^o$, $90^o$ and $135^o$. Energy and entropy were calculated from these matrices to compose a 40-dimensional feature vector \cite{haralickTSMC1973,ChristophPR2004}.

\textit{Gabor filters}: it convolves an image by a bank of Gabor filters (i.e., different scales and orientations). In the experiments, a bank of 40 filters (8 rotations and 5 scales) was used. The energy of each convolved image is used compose the feature vector; in this case a 40-dimensional feature vector. Additional information can be found in \cite{Bianconi2007,Jain1991,gaborJIEE1946}.

\textit{Artificial crawlers}: $N$ artificial crawler, as those explained earlier, are performed in a texture image. Four features vectors are then calculated: (i) the number of live artificial crawlers at each iteration, (ii) the number of settled artificial crawlers at each iteration, (iii) a histogram of the colony size formed by a certain radius and (iv) scale distribution of the colonies. Finally, the four features vectors are concatenated to compose a single vector. A complete description of the original method can be found in \cite{ZhangIAT2004,ZhangIJPRAI2005}.

\textit{Deterministic tourist walk}: this method \cite{backesPR2010} is an agent-based method that builds a joint probability distribution of transient and attractor sizes for different values of memory sizes and two walking rules. In the experiments below, we used memory sizes ranging from $0$ to $5$.

\textit{Multi Fractal Spectrum}: this method \cite{Xu2009} extracts the fractal dimension of three categorization of the image: intensity, energy of edges, and energy of the Laplacian.
For each categorization, a 26-dimensional MFS vector of uniformly spaced values was computed, totaling a feature vector of 78 dimensions.

\textit{Uniform rotation-invariant local binary pattern}: the LBP method \cite{Ojala2002} calculates the co-occurrence of gray-levels in circular neighborhoods. We used three different spatial resolutions $P$ and three different angular resolutions $R$ $-$ $(P,R)$: (8,1), (16,2) and (24,3).

In Table \ref{tab:comparisonBrodatz} we present the comparison of the texture methods on the Brodatz dataset.
The proposed method provided comparable results to the local binary patterns and superior results to the other state-of-the-art methods. 
Though local binary patterns features perform slightly better than ours, the test also indicates that the proposed method significantly improves the success rate over the original artificial crawler, i.e., from 89.75\% to 99.25\%. Despite of the Brodatz dataset is widely used for texture classification, it does not contains textures with changes in terms of lighting conditions and perspectives.

\begin{table}[!htbp]
	\centering
	\caption{The experimental results for texture methods in the Brodatz database.}
	\small
		\begin{tabular}{|c|c|c|}
			\hline
			Method & Correctly classified & Success rate\\
			\hline
			Fourier descriptors & 346 & 86.50 ($\pm 6.58$) \\
			Artificial Crawler & 359 & 89.75 ($\pm 4.76$) \\
			Co-occurrence matrices & 365 & 91.25 ($\pm 2.65$)  \\
			Multi Fractal Spectrum & 373 & 93.25 ($\pm 2.37$) \\
			Gabor filter & 381 & 95.25 ($\pm 3.43$)\\
			Deterministic tourist walk & 382 & 95.50 ($\pm 3.12$)  \\
			Local binary patterns & 399 & 99.75 ($\pm 0.79$) \\
			Proposed method & 397 & 99.25 ($\pm 1.69$) \\
			\hline
		\end{tabular}
		\label{tab:comparisonBrodatz}
\end{table}

To evaluate the methods in textures closer to real-world applications, we also compared the results for the Vistex dataset which are presented in Table \ref{tab:comparisonVistex}.
In this test, our method provided the highest success rate of $95.95\%$, which is superior to the result of the local binary patterns.
Our method significantly improved the success rate compared to the original artificial crawler.
Besides, it can be noted that our method achieved reliable results according to the low standard deviations in both datasets.

\begin{table}[!htbp]
	\centering
	\caption{Experimental results for texture methods in the Vistex database.}
	\small
		\begin{tabular}{|c|c|c|}
			\hline
			Method & Correctly classified & Success rate\\
			\hline
			Fourier descriptors & 672 & 77.78 ($\pm 4.67$) \\
			Artificial Crawler &  691 & 79.98 ($\pm 4.65$) \\
			Co-occurrence matrices &  663 & 76.74 ($\pm 4.91$)  \\
			Deterministic tourist walk & 734 & 84.95 ($\pm 4.13$)  \\
			Multi Fractal Spectrum & 747 & 86.46 ($\pm 3.48$) \\
			Gabor filter & 774 & 89.58 ($\pm 2.61$) \\
			Local binary patterns & 801 & 92.71 ($\pm 2.43$) \\
			Proposed method & 829 & 95.95 ($\pm 2.50$)\\
			\hline
		\end{tabular}
	\label{tab:comparisonVistex}
\end{table}

\section{Conclusion}

In this paper we have proposed a new method based on artificial crawler and fractal dimension for texture classification.
We have demonstrated how the feature vector extraction task can be improved by combining two rules of movement, instead of moving only for the maximum intensity of the neighbor pixels.
Moreover, a strategy using fractal dimension was proposed to characterize the path of movement performed by the artificial crawlers.
The idea of our approach improves the ability of discrimination obtained from the swarm system of artificial crawlers.

Although traditional methods of texture analysis -- e.g. Gabor filters, local binary patterns, and co-occurrence matrices -- have provided satisfactory results, the method proposed here has proved to be superior for characterizing textures on the Vistex dataset. On the Brodatz album, our method achieve the second place, being slightly inferior to the local binary pattern method. 
Experiments on both datasets indicate that our method significantly improved the classification rate with regard to the original artificial crawler method. As future work, we believe that performance gains can be achieved by means of effective descriptors, for example for representing shape.

\section*{Acknowledgements}
W.N.G. acknowledges support from FAPESP (\# 2010/08614-0).
B.B.M. is grateful to FAPESP (\# 2011/02918-0).
O.M.B. gratefully acknowledges the financial support of CNPq (National Council for Scientific and Technological Development, Brazil) (Grant \#308449/2010-0 and \#473893/2010-0) and FAPESP (The State of S\~ao Paulo Research Foundation) (Grant \# 2011/01523-1).







\end{document}